\documentclass[a4paper,10pt]{article}

\usepackage[english]{babel}
\usepackage[utf8]{inputenc}
\usepackage{bera}
\usepackage{inconsolata}

\usepackage[pdftex]{xcolor}
\usepackage[pdftex]{hyperref}
\usepackage[pdftex]{graphicx}
\usepackage{amsmath}
\usepackage{amssymb}
\usepackage{mathrsfs}
\usepackage{subcaption}
\usepackage{array}
\usepackage{listings}
\usepackage{algpseudocode}
\usepackage{tikz}
\usepackage{pgfplots}
\usetikzlibrary{chains,positioning,backgrounds,mindmap,shadows,trees,decorations,arrows,matrix,fit,shapes.geometric,shapes.symbols,lindenmayersystems}

\definecolor{minted}{rgb}{0.95,0.95,1.0}
\definecolor{myblue}{rgb}{0.28,0.24,0.55}
\definecolor{myred}{rgb}{0.71,0.14,0.07}
\definecolor{mygreen}{rgb}{0.41,0.55,0.41}
\definecolor{bg}{RGB}{240,240,225}

\newcommand{\citefig}[1]{(see fig.~\ref{#1}, page~\pageref{#1})}
\newcommand{\citelst}[1]{(see listing~\ref{#1}, p.~\pageref{#1})}

\lstdefinestyle{python}{
	language=Python,
	backgroundcolor=\color{minted},
	basicstyle=\ttfamily\footnotesize,
	keywordstyle=\bfseries\color{green!40!black},
	commentstyle=\itshape\color{purple!40!black},
	identifierstyle=\color{blue},
	stringstyle=\color{orange},
	breaklines=true,
	belowcaptionskip=1\baselineskip,
	xleftmargin=\parindent,
	tabsize=4,
	numbers=left,
	captionpos=b,
	frame=l,
	fontadjust
}

\lstdefinestyle{bash}{
	language=bash,
	basicstyle=\ttfamily\tiny,
	breaklines=true,
	belowcaptionskip=1\baselineskip,
	xleftmargin=\parindent,
	tabsize=4,
	captionpos=b,
	frame=l,
	fontadjust
}

\hypersetup{%
	hypertexnames=false,
	pdfstartview=FitH,
	colorlinks=true,
	urlcolor=myblue,
	linkcolor=myred,
	citecolor=mygreen,
	bookmarksnumbered=true,
	bookmarksopen=true,
	bookmarksopenlevel=2}

\makeindex

\title{Hacking of the AES with Boolean Functions}

\author{Michel~Dubois\thanks{e-mail: michel.dubois@esiea.fr}\\%
	\scriptsize{Operational Cryptology and Virology Laboratory}%
\and \'Eric Filiol\thanks{e-mail: eric.filiol@esiea.fr}\\%
	\scriptsize{Operational Cryptology and Virology Laboratory}}

\begin{document}

\maketitle

\begin{abstract}
One of the major issues of cryptography is the cryptanalysis of cipher algorithms. Cryptanalysis is the study of methods for obtaining the meaning of encrypted information, without access to the secret information that is normally required. Some mechanisms for breaking codes include differential cryptanalysis, advanced statistics and brute-force.

Recent works also attempt to use algebraic tools to reduce the cryptanalysis of a block cipher algorithm to the resolution of a system of quadratic equations describing the ciphering structure. 

In our study, we will also use algebraic tools but in a new way: by using Boolean functions and their properties. A Boolean function is a function from $F_2^n\to F_2$ with $n>1$, characterized by its truth table. The arguments of Boolean functions are binary words of length $n$. Any Boolean function can be represented, uniquely, by its algebraic normal form which is an equation which only contains additions modulo 2 -- the \texttt{XOR} function -- and multiplications modulo 2 -- the \texttt{AND} function.

Our aim is to describe the AES algorithm as a set of Boolean functions then calculate their algebraic normal forms by using the Möbius transforms. After, we use a specific representation for these equations to facilitate their analysis and particularly to try a combinatorial analysis. Through this approach we obtain a new kind of equations system. This equations system is more easily implementable and could open new ways to cryptanalysis.
\end{abstract}


\section{Introduction}

The block cipher algorithms are a family of cipher algorithms which use symmetric key and work on fixed length blocks of data.

Since Novembre 26, 2001, the block cipher algorithm ``Rijndael'', became the successor of DES under the name of ``Advanced Encryption Standard'' (AES). Its designers, Joan Daemen and Vincent Rijmen used algebraic tools to give to their algorithm an unequaled level of assurance against the standard statistical techniques of cryptanalysis. The AES can process data blocks of 128 bits, using cipher keys with lengths of 128, 192, and 256 bits~\cite{fips197}.

One of the major issues of cryptography is the cryptanalysis of cipher algorithms. Cryptanalysis is the study of methods for obtaining the meaning of encrypted information, without access to the secret information that is normally required. Some mechanisms for breaking codes include differential cryptanalysis, advanced statistics and brute-force.

Recent works like \cite{easwaes}, attempt to use algebraic tools to reduce the cryptanalysis of a block cipher algorithm to the resolution of a system of quadratic equations  describing the ciphering structure. As an example, Nicolas Courtois and Josef Pieprzyk have described the AES-128 algorithm as a system of 8000 quadratic equations with 1600 variables \cite{cobcwosoe}. Unfortunately, these approaches are infeasible because of the difficulty of solving large systems of equations.

We will also use algebraic tools but in a new way by using Boolean functions and their properties. Our aim is to describe a block cipher algorithm as a set of Boolean functions then calculate their algebraic normal forms by using the Möbius transforms.

In our study, we will test our approach on the AES algorithm. Our goal is to describe it under the form of systems of Boolean functions and to calculate their algebraic normal forms by using the Möbius transforms. The system of equations obtained is more easily implementable and could open new ways to cryptanalysis of the AES.

\section{Boolean functions}

\subsection{Definition}

Let be the set $B=\{0,1\}$ and $\mathcal{B}_2=\{B, \land, \lor, \lnot\}$ a Boolean algebra, then $\mathcal{B}_2^n = (x_1, x_2, \cdots , x_n)$ such that $x_i \in \mathcal{B}_2$ and $1 \leqq i \leqq n$, is a subset of $\mathcal{B}_2$ containing all $n$-tuples of $0$ and $1$. The variable $x_i$ is called Boolean variable if she only accepts values from $B$, that is to say, if and only if $x_i=0$ or $x_i=1$ regardless of $1 \leqq i \leqq n$.

A Boolean function of degree $n$ with $n>1$ is a function $f$ defined from $\mathcal{B}_2^n \to \mathcal{B}_2$, that is to say built from Boolean variables and agreeing to return values only in the set $B=\{0,1\}$.

For example, the function $f(x_1, x_2)=x_1 \land \lnot x_2$ defined from $\mathcal{B}_2^2 \to \mathcal{B}_2$ is a Boolean function of degree two with:

\begin{align}
	f(0,0) = 0\\
	f(0,1) = 0\\
	f(1,0) = 1\\
	f(1,1) = 0
\end{align}

Let $n$ and $m$ be two positive integers. A vector Boolean function is a Boolean function $f$ defined from $\mathcal{B}_2^n \to \mathcal{B}_2^m$.

An S-box is a vector Boolean function.

Finally, we can define a random Boolean function as a Boolean function $f$ whose values are independent and identically distributed random variables, that is to say: $$\forall (x_1, x_2, \cdots , x_n) \in \mathcal{B}_2^n,\quad P[f(x_1, x_2, \cdots , x_n) = 0] = \frac{1}{2}$$

The number of Boolean functions is limited and depends on $n$. Thus, there is $2^{2^n}$ Boolean functions. Similarly, the number of vector Boolean functions is limited and depends on $n$ and $m$. Thus, there exists $\left(2^m\right)^{2^n}$ vector Boolean functions.

If we take, for example, $n=2$ then there exists $\left(2^2\right)^2 = 16$ Boolean functions of degree two. These 16 Boolean functions are presented in the table in figure~\ref{fig:16bool} page~\pageref{fig:16bool}. Among the Boolean functions of degree 2, the best known are the functions OR, AND and XOR \citefig{fig:or}, \citefig{fig:and} and \citefig{fig:xor}.

\begin{figure}
	\begin{center}
	\begin{tabular}{| c | c |}
		\hline
$f_0$ & $0$ \tabularnewline \hline
$f_1$ & $x_1 \land x_2$ \tabularnewline \hline
$f_2$ & $x_1 \land \lnot x_2$ \tabularnewline \hline
$f_3$ & $x_1$ \tabularnewline \hline
$f_4$ & $\lnot x_1 \land x_2$ \tabularnewline \hline
$f_5$ & $x_2$ \tabularnewline \hline
$f_6$ & $x_1 \veebar x_2$ \tabularnewline \hline
$f_7$ & $x_1 \lor x_2$ \tabularnewline \hline
$f_8$ & $\lnot(x_1 \lor x_2)$ \tabularnewline \hline
$f_9$ & $\lnot(x_1 \veebar x_2)$ \tabularnewline \hline
$f_{10}$ & $\lnot x_2$ \tabularnewline \hline
$f_{11}$ & $x_1 \lor \lnot x_2$ \tabularnewline \hline
$f_{12}$ & $\lnot x_1$ \tabularnewline \hline
$f_{13}$ & $\lnot x_1 \lor x_2$ \tabularnewline \hline
$f_{14}$ & $\lnot(x_1 \land x_2)$ \tabularnewline \hline
$f_{15}$ & $1$ \tabularnewline \hline
	\end{tabular}
	\end{center}
	\caption{The 16 Boolean functions of degree 2}
	\label{fig:16bool}
\end{figure}

\begin{figure}
	\begin{center}
		\includegraphics[width=0.90\textwidth]{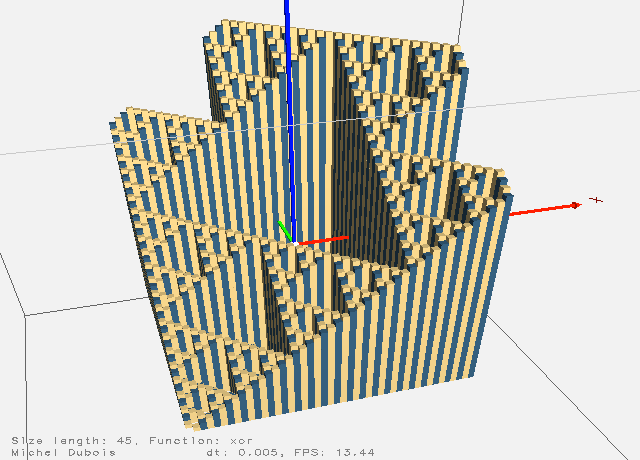}
	\end{center}
	\caption{The \texttt{XOR} function}
	\label{fig:xor}
\end{figure}

\begin{figure}
	\begin{center}
		\includegraphics[width=0.90\textwidth]{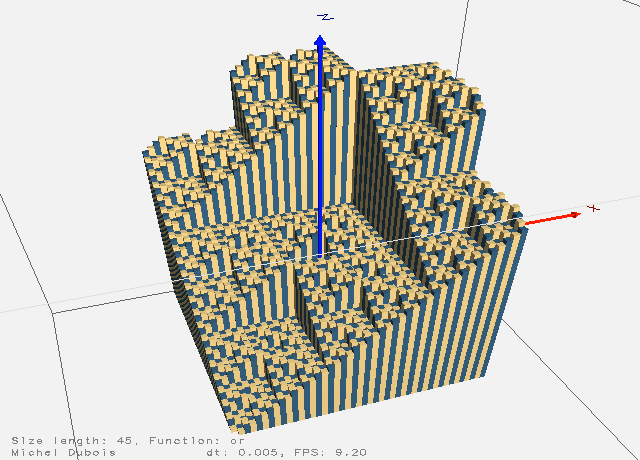}
	\end{center}
	\caption{The \texttt{OR} function}
	\label{fig:or}
\end{figure}

\begin{figure}
	\begin{center}
		\includegraphics[width=0.90\textwidth]{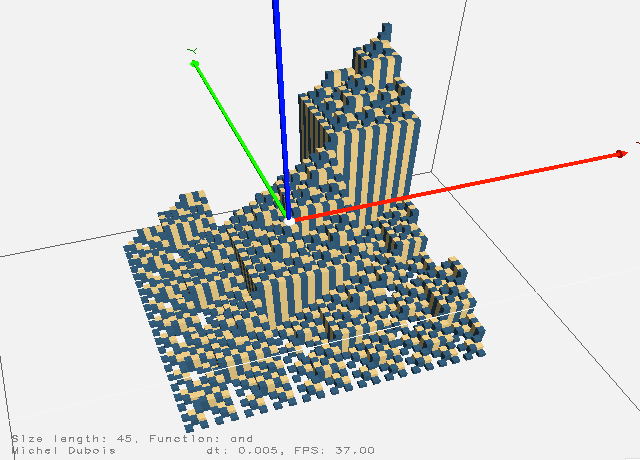}
	\end{center}
	\caption{The \texttt{AND} function}
	\label{fig:and}
\end{figure}

The support $supp(f)$ of a Boolean function is the set of elements $x$ such that $f(x)\neq0$, the Hamming weight $wt(f)$ of a Boolean function is the cardinal from its support and we have: $$ wt(f) = | \{x \in \mathcal{B}_2^n \quad | \quad f(x)=1 \} |$$

A Boolean function is called balanced if $wt(f) = 2^{n-1}$. Similarly, a Boolean vector function $\mathcal{B}_2^n \to \mathcal{B}_2^m$ is said to be balanced if $wt(f) = 2^{n-m}$~\cite{vbffc}.

For example, the support of the function $f(x_1, x_2) = x_1 \lor x_2$, corresponding to logical \texttt{OR} is $supp(f) = \{(0,1), (1,0), (1,1)\}$ and its weight is $wt(\text{f}) = 3$.

\subsection{Representations}

There are multiple representations of Boolean functions. We'll look at the most common -- the truth table -- and that we will use later -- a representation in $GF(2)$.

\subsubsection{The truth table}

The different values taken by a Boolean function may be presented in the form of a table called truth table. The truth table characterizes a Boolean function.

For example, the truth table of the Boolean function of degree four $$f(x_1, x_2, x_3, x_4)=(x_1 \land x_2) \lor (x_3 \land x_4)$$ is presented in figure~\ref{fig:boole_f} page~\pageref{fig:boole_f}.

\begin{figure}
	\begin{center}
	\begin{footnotesize}
	\begin{tabular}{| c | c | c | c | c |}
		\hline
		$x_1$ & $x_2$ & $x_3$ & $x_4$ & $f(x_1, x_2, x_3, x_4)$ \tabularnewline \hline
		$0$ & $0$ & $0$ & $0$ & $0$ \tabularnewline \hline
		$0$ & $0$ & $0$ & $1$ & $0$ \tabularnewline \hline
		$0$ & $0$ & $1$ & $0$ & $0$ \tabularnewline \hline
		$0$ & $0$ & $1$ & $1$ & $1$ \tabularnewline \hline
		$0$ & $1$ & $0$ & $0$ & $0$ \tabularnewline \hline
		$0$ & $1$ & $0$ & $1$ & $0$ \tabularnewline \hline
		$0$ & $1$ & $1$ & $0$ & $0$ \tabularnewline \hline
		$0$ & $1$ & $1$ & $1$ & $1$ \tabularnewline \hline
		$1$ & $0$ & $0$ & $0$ & $0$ \tabularnewline \hline
		$1$ & $0$ & $0$ & $1$ & $0$ \tabularnewline \hline
		$1$ & $0$ & $1$ & $0$ & $0$ \tabularnewline \hline
		$1$ & $0$ & $1$ & $1$ & $1$ \tabularnewline \hline
		$1$ & $1$ & $0$ & $0$ & $1$ \tabularnewline \hline
		$1$ & $1$ & $0$ & $1$ & $1$ \tabularnewline \hline
		$1$ & $1$ & $1$ & $0$ & $1$ \tabularnewline \hline
		$1$ & $1$ & $1$ & $1$ & $1$ \tabularnewline \hline
	\end{tabular}
	\end{footnotesize}
	\end{center}
\caption{Truth table of the Boolean function $f(x_1, x_2, x_3, x_4)=(x_1 \land x_2) \lor (x_3 \land x_4)$}
	\label{fig:boole_f}
\end{figure}

Similarly, the table in figure~\ref{fig:16bool_tt} page~\pageref{fig:16bool_tt} details the truth tables of the 16 Boolean functions of degree two.

\begin{figure}
	\begin{center}
	\begin{footnotesize}
	\begin{tabular}{| c | c | c | c | c | c | c | c | c | c |}
		\hline
$x_1$ & $x_2$ & $f_0$ & $f_1$ & $f_2$ & $f_3$ & $f_4$ & $f_5$ & $f_6$ & $f_7$ \tabularnewline \hline
0 & 0 & 0 & 0 & 0 & 0 & 0 & 0 & 0 & 0  \tabularnewline \hline
0 & 1 & 0 & 0 & 0 & 0 & 1 & 1 & 1 & 1 \tabularnewline \hline
1 & 0 & 0 & 0 & 1 & 1 & 0 & 0 & 1 & 1 \tabularnewline \hline
1 & 1 & 0 & 1 & 0 & 1 & 0 & 1 & 0 & 1 \tabularnewline \hline
	\end{tabular}

	\bigskip

	\begin{tabular}{| c | c | c | c | c | c | c | c | c | c |}
		\hline
$x_1$ & $x_2$ & $f_8$ & $f_9$ & $f_{10}$ & $f_{11}$ & $f_{12}$ & $f_{13}$ & $f_{14}$ & $f_{15}$ \tabularnewline \hline
0 & 0 & 1 & 1 & 1 & 1 & 1 & 1 & 1 & 1 \tabularnewline \hline
0 & 1 & 0 & 0 & 0 & 0 & 1 & 1 & 1 & 1 \tabularnewline \hline
1 & 0 & 0 & 0 & 1 & 1 & 0 & 0 & 1 & 1 \tabularnewline \hline
1 & 1 & 0 & 1 & 0 & 1 & 0 & 1 & 0 & 1 \tabularnewline \hline
	\end{tabular}
	\end{footnotesize}
	\end{center}
	\caption{The truth tables of the 16 Boolean functions of degree 2}
	\label{fig:16bool_tt}
\end{figure}

\subsubsection{Representation in GF(2)}

A Boolean function can also be presented in the form of a series of conjunctions including disjunctions, negations and/or variables. This is called the conjunctive normal form. Thus, the sequence $f = (a \lor b) \land (\lnot a \lor b)$ is the conjunctive normal form of the $f$ function. Conversely, a Boolean function can be presented in the form of a series of disjunctions including conjunctions, negations and/or variables. This is called the disjunctive normal form. Thus, the sequence $g = (a \land b) \lor (\lnot a \land b)$ is the disjunctive normal form of the function $g$.

Now let the representation of Boolean functions in $GF(2)$.

\begin{figure}
\begin{center}
\begin{tikzpicture}
	\tikzstyle{dark} = [draw=blue!80, fill=blue!20, minimum width=18pt, minimum height=18pt, font=\bfseries]
	\tikzstyle{light} = [draw=black!80, fill=black!20, minimum width=18pt, minimum height=18pt]

	\matrix[column sep=1pt, row sep=1pt] (and) {
		\node[dark] {$\land$}; & \node[dark] {$0$}; & \node[dark] {$1$}; \\
		\node[dark] {$0$}; & \node[light] {$0$}; & \node[light] {$0$}; \\
		\node[dark] {$1$}; & \node[light] {$0$}; & \node[light] {$1$}; \\
	};

	\matrix[column sep=1pt, row sep=1pt, right=8pt of and] (or) {
		\node[dark] {$\lor$}; & \node[dark] {$0$}; & \node[dark] {$1$}; \\
		\node[dark] {$0$}; & \node[light] {$0$}; & \node[light] {$1$}; \\
		\node[dark] {$1$}; & \node[light] {$1$}; & \node[light] {$1$}; \\
	};

	\matrix[column sep=1pt, row sep=1pt, right=8pt of or] (not) {
		\node[dark] {$a$}; & \node[dark] {$0$}; & \node[dark] {$1$}; \\
		\node[dark] {$\lnot a$}; & \node[light] {$1$}; & \node[light] {$0$}; \\
	};
\end{tikzpicture}
\end{center}
\caption{Rules for Boolean algebra with two elements}
\label{fig:alg_boole_2}
\end{figure}

The set $B=\{0, 1\}$ associated with $\land$, $\lor$ and $\lnot$ operations is the Boolean algebra $\mathcal{B}_2=\{B, \land, \lor, \lnot\}$ with the truth tables of the operations described in figure~\ref{fig:alg_boole_2} page~\pageref{fig:alg_boole_2}. If we introduce the two binary operations $\oplus$ and $\bullet$ defined by the truth tables in figure~\ref{fig:gf2} page~\pageref{fig:gf2}, then $\mathcal{B}_2$ and the Galois field $GF(2)$ are similar. More specifically, the Boolean algebra $(B, \land, \lor, \lnot)$ and the field $(GF(2), \bullet, \oplus)$ are related by the following transformation formulas:

\begin{tabular}{l l}
	$a \land  b = a \bullet b$ & $a \bullet b = a \land b$ \tabularnewline
	$a \lor  b = a \oplus b \oplus (a \bullet b)$ & $a \oplus b = (a \land \lnot b) \lor ( \lnot a \land b )$ \tabularnewline
	$\lnot a = a \oplus 1$ & \tabularnewline
\end{tabular}

\begin{figure}
\begin{center}
\begin{tikzpicture}
	\tikzstyle{dark} = [draw=blue!80, fill=blue!20, minimum width=18pt, minimum height=18pt, font=\bfseries]
	\tikzstyle{light} = [draw=black!80, fill=black!20, minimum width=18pt, minimum height=18pt]

	\matrix[column sep=1pt, row sep=1pt] (and) {
		\node[dark] {$\bullet$}; & \node[dark] {$0$}; & \node[dark] {$1$}; \\
		\node[dark] {$0$}; & \node[light] {$0$}; & \node[light] {$0$}; \\
		\node[dark] {$1$}; & \node[light] {$0$}; & \node[light] {$1$}; \\
	};

	\matrix[column sep=1pt, row sep=1pt, right=8pt of and] (or) {
		\node[dark] {$\oplus$}; & \node[dark] {$0$}; & \node[dark] {$1$}; \\
		\node[dark] {$0$}; & \node[light] {$0$}; & \node[light] {$1$}; \\
		\node[dark] {$1$}; & \node[light] {$1$}; & \node[light] {$0$}; \\
	};
\end{tikzpicture}
\end{center}
\caption{Truth tables of $\bullet$ and $\oplus$}
\label{fig:gf2}
\end{figure}

We can now define a Boolean function as a function $f:\mathbb{F}_2^n \to \mathbb{F}_2$ with $\mathbb{F}_2^n$ the set of binary vectors of length $n>1$. The Hamming weight $wH(x)$ of the binary vector $x \in \mathbb{F}_2^n$ is the number of non-zero coordinates, that is to say the size of the set $\{ i \in \mathbb{N} \quad | \quad x_i \ne 0 \}$. The Hamming weight of a Boolean function $f:\mathbb{F}_2^n \to \mathbb{F}_2$ is the size of its support. Finally, the Hamming distance between two Boolean functions $f$ and $g$ is the size of the set $\{ x \in \mathbb{F}_2^n \quad | \quad f(x) \ne g(x) \}$.

Among the classic representation of Boolean functions, the most frequently used in cryptography is the polynomial representation in $n$-variable on $GF(2)$. This representation is of the form \cite{bfcecc}:

\begin{align}
	f(x)  & = \bigoplus_{I \in P(N)} a_I \left( \prod_{i \in I} x_i \right) \notag \\
	& = \bigoplus_{I \in P(N)} a_i x^I \notag
\end{align}

$P(N)$ denotes the set of powers of $N = \{ 1, \cdots, n \}$. Each coordinate $x_i$ appears in this polynomial with an exponent equal to at least one, because in $\mathbb{F}_2$ we have$x^2 = x$. This representation is described in $\mathbb{F}_2 [x_1,\cdots,x_n] / (x_1^2 \oplus x_1, \cdots, x_n^2 \oplus x_n)$.

This representation of Boolean functions in $GF(2)$ is called Reed-Muller expansion or polynomials of Zhegalkin (\cite{abf} page 169) or, more commonly, \textsl{algebraic normal form} (ANF). The degree of $ANF(f)$ is the highest degree of monomials of $ANF(f)$ with non-zero coefficients. Finally, the algebraic normal form of a Boolean function exists and is unique.

In summary, any Boolean function can be represented uniquely by its algebraic normal form as the equation:

\begin{align}
	f(x_1,\cdots,x_n) & = a_0 +\notag \\
	& a_1x_1 + a_2x_2 + \cdots + a_nx_n +\notag \\
	& a_{1,2}x_1x_2 + \cdots + a_{n-1,n}x_{n-1}x_n +\notag \\
	& \cdots +\notag \\
	& a_{1,2,\ldots,n}x_1x_2\ldots x_n \notag
\end{align}

Consider an example. Let the function $f$ described by the following truth table:

\begin{center}
\begin{tabular}{| c c c | c |}
	\hline
	$x_1$ & $x_2$ & $x_3$ & $f(x)$ \\
	\hline
	$0$ & $0$ & $0$ & $0$ \\
	$0$ & $0$ & $1$ & $1$ \\
	$0$ & $1$ & $0$ & $0$ \\
	$0$ & $1$ & $1$ & $0$ \\
	$1$ & $0$ & $0$ & $0$ \\
	$1$ & $0$ & $1$ & $1$ \\
	$1$ & $1$ & $0$ & $0$ \\
	$1$ & $1$ & $1$ & $1$ \\
	\hline
\end{tabular}
\end{center}

The weight of the function $f$ is $wt(f) = 3$. So we can reduce $f$ to the sum of 3 atomic functions $f_1$, $f_2$ and $f_3$. The function $f_1=1$ if and only if $1\oplus x_1 = 1$, $1\oplus x_2 = 1$ and $x_3 = 1$. From this we can deduce that the ANF of the function $f_1$ can be obtained by expanding the product $(1\oplus x_1)(1\oplus x_2)x_3$. Applying this reasoning to the functions $f_2$ and $f_3$ we get the following equation:

\begin{align}
	ANF(f) & = (1\oplus x_1)(1\oplus x_2)x_3 \oplus x_1(1\oplus x_2)x_3 \oplus x_1x_2x_3 \notag \\
	& = x_1x_2x_3 \oplus x_1x_3 \oplus x_3
\end{align}

\section{Mechanism of the equations}

After this brief presentation of Boolean functions, we have the necessary tools for the development of systems of Boolean equations describing the \textsl{Advanced Encryption standard}.

\subsection{Möbius transform}

We have just seen how to generate normal algebraic form (ANF) of a Boolean function. The presented method is not easily automatable in a computer program. So we will prefer the use of the Möbius transform.

The Möbius transform of the Boolean function $f$ is defined by~\cite{mccarty_book}:

\begin{align}
	TM(f) &:\mathbb{F}_2^n \to \mathbb{F}_2 \notag \\
	& u = \bigoplus_{v\leqslant u}f(v) \text{mod} 2\notag
\end{align}

with $v\leqslant u$ if and only if $\forall i, v_i=1\Rightarrow u_i=1$.

From there, we can define the normal algebraic form of a Boolean function $f$ in $n$ variables:

$$\bigoplus_{u=(u_1,\cdots,u_n)\in\mathbb{F}_2^n} TM(u)x_1^{u_1}\cdots x_n^{u_n}$$

To better understand the mechanisms involved in the use of the Möbius transform, take an example with the \texttt{MajParmi3}. This function from $\mathbb{F}_2^3\to\mathbb{F}_2$ is characterized by the truth table shown in figure~\ref{fig:majparmi3} page~\pageref{fig:majparmi3}.

\begin{figure}
	\begin{center}
	\begin{tabular}{ >{\centering}p{18pt} | >{\centering}p{18pt} | >{\centering}p{18pt} | >{\centering}p{52pt} }
		$x_1$ & $x_2$ & $x_3$ & \texttt{MajParmi3} \tabularnewline
		\hline
		$0$ & $0$ & $0$ & $0$ \tabularnewline
		$0$ & $0$ & $1$ & $0$ \tabularnewline
		$0$ & $1$ & $0$ & $0$ \tabularnewline
		$0$ & $1$ & $1$ & $1$ \tabularnewline
		$1$ & $0$ & $0$ & $0$ \tabularnewline
		$1$ & $0$ & $1$ & $1$ \tabularnewline
		$1$ & $1$ & $0$ & $1$ \tabularnewline
		$1$ & $1$ & $1$ & $1$ \tabularnewline
	\end{tabular}
	\end{center}
	\caption{The truth table of the function \texttt{MajParmi3}}
	\label{fig:majparmi3}
\end{figure}

Calculating the Möbius transform of the function we get the result of figure~\ref{fig:rm-mp3} page~\pageref{fig:rm-mp3}.

\begin{figure}
	\begin{center}
	\begin{tabular}{ >{\centering}p{10pt} | >{\centering}p{10pt} | >{\centering}p{10pt} | >{\centering}p{52pt} >{\centering}p{10pt} >{\centering}p{10pt} | >{\centering}p{10pt} | >{\centering}p{10pt} | >{\centering}p{10pt} | >{\centering}p{30pt} |}
		$x_1$ & $x_2$ & $x_3$ & \texttt{MajParmi3} & $\to$ & \multicolumn{4}{c | }{compute of $TM(f)$} & $TM(f)$ \tabularnewline
		\hline
		$0$ & $0$ & $0$ & $0$ & $\to$ & $0$ & $0$ & $0$& $0$ & $0$ \tabularnewline
		\cline{9-9}
		$0$ & $0$ & $1$ & $0$ & $\to$ & $0$ & $0$ & $0$& $0$ & $0$ \tabularnewline
		\cline{8-9}
		$0$ & $1$ & $0$ & $0$ & $\to$ & $0$ & $0$ & $0$& $0$ & $0$ \tabularnewline
		\cline{9-9}
		$0$ & $1$ & $1$ & $1$ & $\to$ & $1$ & $1$ & $1$& $1$ & $1$ \tabularnewline
		\cline{6-9}
		$1$ & $0$ & $0$ & $0$ & $\to$ & $0$ & $0$ & $0$& $0$ & $0$ \tabularnewline
		\cline{9-9}
		$1$ & $0$ & $1$ & $1$ & $\to$ & $1$ & $1$ & $1$& $1$ & $1$ \tabularnewline
		\cline{8-9}
		$1$ & $1$ & $0$ & $1$ & $\to$ & $1$ & $1$ & $1$& $1$ & $1$ \tabularnewline
		\cline{9-9}
		$1$ & $1$ & $1$ & $1$ & $\to$ & $1$ & $0$ & $1$& $0$ & $0$ \tabularnewline
	\end{tabular}
	\end{center}
	\caption{Calculating the Möbius transform for \texttt{MajParmi3}}
	\label{fig:rm-mp3}
\end{figure}

After the Möbius transform of the function obtained, we take the $\mathbb{F}_2^3$ for which $TM(\text{MajParmi3}) \neq 0$. In our case we have the triplets $(0,1,1)$, $(1,0,1)$, $(1,1,0)$ from which we can deduce the equation:

$$\text{MajParmi3}(x_1, x_2, x_3) = x_2x_3 + x_1x_3 + x_1x_2$$

With the addition corresponding to a \texttt{XOR} and multiplication to a \texttt{AND}.

The implementation of the Möbius transform in Python is performed by the two functions described in the listing~\ref{lst:tm} page~\pageref{lst:tm}.

\begin{lstlisting}[style=python, caption={Calculation of the Möbius transform in python},label=lst:tm]
def xorTab(t1, t2):
	"""Takes two tabs t1 and t2 of same lengths and returns t1 XOR t2."""
	result = ''
	for i in xrange(len(t1)):
		result += str(int(t1[i]) ^ int(t2[i]))
	return result

def moebiusTransform(tab):
	"""Takes a tab and return tab[0 : len(tab)/2],
	tab[0 : len(tab)/2] ^ tab[len(tab)/2 : len(tab)].
	usage: moebiusTransform(1010011101010100) --> [1100101110001010]"""
	if len(tab) == 1:
		return tab
	else:
		t1 = tab[0 : len(tab)/2]
		t2 = tab[len(tab)/2 : len(tab)]
		t2 = xorTab(t1, t2)
		t1 = moebiusTransform(t1)
		t2 = moebiusTransform(t2)
		t1 += t2
		return t1
\end{lstlisting}

\subsection{Formatting equations}

To facilitate the analysis and in particular to try a combinatorial study we will implement a specific presentation for equations thus obtained.

The AES algorithm takes 128 bits as input and provides 128 bits as output. So we will have Boolean functions $F_2^{128}\to F_2^{128}$. The guiding principle is to generate a file by bit, we will have at the end 128 files. Each file containing the Boolean equation of the concerned bit.

In each file, the Boolean equation is presented under the form of lines containing sequences of 0 and 1. Each line describes a monomial of the equation and the transition from one line to another means applying a \texttt{XOR}.

In order to facilitate understanding of the chosen mechanism we describe the realization of file corresponding to one bit $b_1$ from his equation to the file formalism in figure~\ref{fig:fichiers_bit} page~\pageref{fig:fichiers_bit}.

\begin{figure}
\footnotesize{%
\begin{center}
	\begin{align}
		f(b_1) = 1 \oplus b_{15}b_{16} \oplus b_{14} \oplus b_{14}b_{16} \oplus b_{13} \oplus b_{13}b_{15}\oplus b_{13}b_{15}b_{16} \notag\\
		\oplus b_{4} \oplus b_{3}b_{4} \oplus b_{2}b_{4} \oplus b_{2}b_{3} \oplus b_{2}b_{3}b_{4} \oplus b_{1}b_{3} \notag\\ 
		\oplus b_{1}b_{3}b_{4} \oplus b_{1}b_{2} \oplus b_{1}b_{2}b_{3}\notag
	\end{align}

	\vspace{10pt}

	\begin{tabular}{ p{45pt} | >{\centering}p{8pt} >{\centering}p{100pt} | }
		\cline{2-3}
		& &\tabularnewline
		$1$ & 1 & 0000000000000000 \tabularnewline
		$b_{15}b_{16}$ & 0 & 0000000000000011 \tabularnewline
		$b_{14}$ & 0 & 0000000000000100 \tabularnewline
		$b_{14}b_{16}$ & 0 & 0000000000000101 \tabularnewline
		$b_{13}$ & 0 & 0000000000001000 \tabularnewline
		$b_{13}b_{15}$ & 0 & 0000000000001010 \tabularnewline
		$b_{13}b_{15}b_{16}$ & 0 & 0000000000001011 \tabularnewline
		$b_{4}$ & 0 & 0001000000000000 \tabularnewline
		$b_{3}b_{4}$ & 0 & 0011000000000000 \tabularnewline
		$b_{2}b_{4}$ & 0 & 0101000000000000 \tabularnewline
		$b_{2}b_{3}$ & 0 & 0110000000000000 \tabularnewline
		$b_{2}b_{3}b_{4}$ & 0 & 0111000000000000 \tabularnewline
		$b_{1}b_{3}$ & 0 & 1010000000000000 \tabularnewline
		$b_{1}b_{3}b_{4}$ & 0 & 1011000000000000 \tabularnewline
		$b_{1}b_{2}$ & 0 & 1100000000000000 \tabularnewline
		$b_{1}b_{2}b_{3}$ & 0 & 1110000000000000 \tabularnewline
		\cline{2-3}
	\end{tabular}
\end{center}
}
\caption{File for the bit $b_1$}
\label{fig:fichiers_bit}
\end{figure}

\section{Application to AES}

\subsection{The equations for AES}

We will now apply to the AES the mechanism described above. The difficulty with our approach is that the encryption functions of the AES algorithm takes 128 bits as input and provides 128 bits as output. So we will have Boolean functions $F_2^{128}\to F_2^{128}$ and it is impossible to calculate their truth tables. Indeed, in this case, we have $2^{128}=3,402823 \times 10^{38}$ possible combinations of 128-bit blocks and the space storage needed to archive these blocks is $3,868562 \times 10^{25}$~terabytes.

So we have to find a way to describe the AES encryption functions in the form of Boolean functions without using their truth table.

\subsection{The equations for ciphering functions}

We will now detail the solution implemented for each of the sub-functions of the AES encryption algorithm.

\subsubsection{Solution for SubBytes function}

The function \texttt{SubBytes} is a non-linear substitution that works on every byte of the states array using a substitution table (S-Box).

This function is applied independently to each byte of the input block. So, the S-box of the AES is a function taking 8 bits as input and providing 8-bit as output. So we can describe it as a Boolean function $F_2^8\to F_2^8$. From there, we can calculate the truth table of the S-Box and use the Möbius transform for obtain the normal algebraic form of the S-Box. Then applying the results to the 16 bytes of input block, we get 128 equations, each describing a block bit.

For example, the equation of the processing of the bit $b_{127}$ by the function \texttt{SubByte} is given in figure \ref{fig:bitsubbyte} page~\pageref{fig:bitsubbyte}.

\begin{figure}
	\begin{flushleft}
	\footnotesize{%
	$1 \oplus x_{127} \oplus x_{126}x_{127} \oplus x_{125} \oplus x_{125}x_{126} \oplus x_{124} \oplus x_{124}x_{126} \oplus x_{124}x_{125} \oplus x_{124}x_{125}x_{126} \oplus x_{124}x_{125}x_{126}x_{127} \oplus x_{123} \oplus x_{123}x_{127} \oplus x_{123}x_{126} \oplus x_{123}x_{126}x_{127} \oplus x_{123}x_{125} \oplus x_{123}x_{125}x_{127} \oplus x_{123}x_{125}x_{126} \oplus x_{123}x_{124}x_{127} \oplus x_{123}x_{124}x_{126} \oplus x_{123}x_{124}x_{125}x_{126}x_{127} \oplus x_{122}x_{127} \oplus x_{122}x_{125}x_{127} \oplus x_{122}x_{125}x_{126}x_{127} \oplus x_{122}x_{124}x_{127} \oplus x_{122}x_{124}x_{125} \oplus x_{122}x_{124}x_{125}x_{127} \oplus x_{122}x_{124}x_{125}x_{126} \oplus x_{122}x_{123}x_{126}x_{127} \oplus x_{122}x_{123}x_{125}x_{127} \oplus x_{122}x_{123}x_{125}x_{126}x_{127} \oplus x_{122}x_{123}x_{124}x_{125}x_{127} \oplus x_{121}x_{127} \oplus x_{121}x_{126} \oplus x_{121}x_{126}x_{127} \oplus x_{121}x_{125} \oplus x_{121}x_{125}x_{127} \oplus x_{121}x_{125}x_{126} \oplus x_{121}x_{125}x_{126}x_{127} \oplus x_{121}x_{124}x_{127} \oplus x_{121}x_{124}x_{125}x_{126} \oplus x_{121}x_{124}x_{125}x_{126}x_{127} \oplus x_{121}x_{123} \oplus x_{121}x_{123}x_{127} \oplus x_{121}x_{123}x_{126} \oplus x_{121}x_{123}x_{125}x_{126} \oplus x_{121}x_{123}x_{124}x_{127} \oplus x_{121}x_{123}x_{124}x_{126} \oplus x_{121}x_{123}x_{124}x_{126}x_{127} \oplus x_{121}x_{123}x_{124}x_{125}x_{127} \oplus x_{121}x_{123}x_{124}x_{125}x_{126}x_{127} \oplus x_{121}x_{122} \oplus x_{121}x_{122}x_{126} \oplus x_{121}x_{122}x_{125} \oplus x_{121}x_{122}x_{125}x_{127} \oplus x_{121}x_{122}x_{124}x_{127} \oplus x_{121}x_{122}x_{124}x_{126}x_{127} \oplus x_{121}x_{122}x_{124}x_{125}x_{126} \oplus x_{121}x_{122}x_{123} \oplus x_{121}x_{122}x_{123}x_{127} \oplus x_{121}x_{122}x_{123}x_{126} \oplus x_{121}x_{122}x_{123}x_{126}x_{127} \oplus x_{121}x_{122}x_{123}x_{125} \oplus x_{121}x_{122}x_{123}x_{125}x_{126} \oplus x_{121}x_{122}x_{123}x_{124}x_{127} \oplus x_{121}x_{122}x_{123}x_{124}x_{125} \oplus x_{121}x_{122}x_{123}x_{124}x_{125}x_{127} \oplus x_{120}x_{126}x_{127} \oplus x_{120}x_{125} \oplus x_{120}x_{125}x_{127} \oplus x_{120}x_{125}x_{126}x_{127} \oplus x_{120}x_{124}x_{126} \oplus x_{120}x_{124}x_{125} \oplus x_{120}x_{124}x_{125}x_{127} \oplus x_{120}x_{124}x_{125}x_{126} \oplus x_{120}x_{124}x_{125}x_{126}x_{127} \oplus x_{120}x_{123}x_{127} \oplus x_{120}x_{123}x_{126}x_{127} \oplus x_{120}x_{123}x_{125} \oplus x_{120}x_{123}x_{125}x_{127} \oplus x_{120}x_{123}x_{125}x_{126} \oplus x_{120}x_{123}x_{125}x_{126}x_{127} \oplus x_{120}x_{123}x_{124} \oplus x_{120}x_{123}x_{124}x_{125}x_{126}x_{127} \oplus x_{120}x_{122} \oplus x_{120}x_{122}x_{125} \oplus x_{120}x_{122}x_{125}x_{127} \oplus x_{120}x_{122}x_{125}x_{126}x_{127} \oplus x_{120}x_{122}x_{124} \oplus x_{120}x_{122}x_{124}x_{126}x_{127} \oplus x_{120}x_{122}x_{124}x_{125} \oplus x_{120}x_{122}x_{124}x_{125}x_{126} \oplus x_{120}x_{122}x_{124}x_{125}x_{126}x_{127} \oplus x_{120}x_{122}x_{123}x_{127} \oplus x_{120}x_{122}x_{123}x_{126} \oplus x_{120}x_{122}x_{123}x_{125} \oplus x_{120}x_{122}x_{123}x_{125}x_{127} \oplus x_{120}x_{122}x_{123}x_{125}x_{126}x_{127} \oplus x_{120}x_{122}x_{123}x_{124}x_{127} \oplus x_{120}x_{122}x_{123}x_{124}x_{126} \oplus x_{120}x_{122}x_{123}x_{124}x_{125} \oplus x_{120}x_{122}x_{123}x_{124}x_{125}x_{127} \oplus x_{120}x_{121} \oplus x_{120}x_{121}x_{125} \oplus x_{120}x_{121}x_{125}x_{126} \oplus x_{120}x_{121}x_{125}x_{126}x_{127} \oplus x_{120}x_{121}x_{124} \oplus x_{120}x_{121}x_{124}x_{126} \oplus x_{120}x_{121}x_{124}x_{126}x_{127} \oplus x_{120}x_{121}x_{124}x_{125} \oplus x_{120}x_{121}x_{124}x_{125}x_{126}x_{127} \oplus x_{120}x_{121}x_{123}x_{127} \oplus x_{120}x_{121}x_{123}x_{125}x_{127} \oplus x_{120}x_{121}x_{123}x_{125}x_{126} \oplus x_{120}x_{121}x_{123}x_{124}x_{127} \oplus x_{120}x_{121}x_{123}x_{124}x_{126}x_{127} \oplus x_{120}x_{121}x_{123}x_{124}x_{125}x_{126} \oplus x_{120}x_{121}x_{123}x_{124}x_{125}x_{126}x_{127} \oplus x_{120}x_{121}x_{122} \oplus x_{120}x_{121}x_{122}x_{126} \oplus x_{120}x_{121}x_{122}x_{126}x_{127} \oplus x_{120}x_{121}x_{122}x_{125}x_{126} \oplus x_{120}x_{121}x_{122}x_{125}x_{126}x_{127} \oplus x_{120}x_{121}x_{122}x_{124} \oplus x_{120}x_{121}x_{122}x_{124}x_{127} \oplus x_{120}x_{121}x_{122}x_{124}x_{125}x_{127} \oplus x_{120}x_{121}x_{122}x_{123}x_{126}x_{127} \oplus x_{120}x_{121}x_{122}x_{123}x_{125} \oplus x_{120}x_{121}x_{122}x_{123}x_{125}x_{126} \oplus x_{120}x_{121}x_{122}x_{123}x_{125}x_{126}x_{127} \oplus x_{120}x_{121}x_{122}x_{123}x_{124}x_{126} \oplus x_{120}x_{121}x_{122}x_{123}x_{124}x_{125} \oplus x_{120}x_{121}x_{122}x_{123}x_{124}x_{125}x_{127}$
	}
	\end{flushleft}
	\caption{Equation of the bit $b_{127}$ from fonction \texttt{SubByte}}
	\label{fig:bitsubbyte}
\end{figure}

\subsubsection{Solution for ShiftRows function}

In the \texttt{ShiftRows} function, the bytes of the third column of the state table are shifted cyclically in an offset whose size is dependent on the line number. The bytes of the first line do not suffer this offset.

For this function, we do not need to calculate specific Boolean function. Indeed, the only change made consists to shift bytes in the states array. In our files, this transformation can be easily solved by using a \texttt{XOR}.

Thus, for example, the second byte of the status table becomes the sixth byte after the application of \texttt{ShiftRows}. This results in the following lines:

\begin{center}
\fontsize{5}{6}\selectfont
\begin{ttfamily}
\textcolor{red}{00000000}00000000\textcolor{red}{00000000}00000000\textcolor{red}{00000000}\textbf{10000000}\textcolor{red}{00000000}00000000\textcolor{red}{00000000}00000000\textcolor{red}{00000000}00000000\textcolor{red}{00000000}00000000\textcolor{red}{00000000}00000000\newline\noindent
\textcolor{red}{00000000}00000000\textcolor{red}{00000000}00000000\textcolor{red}{00000000}\textbf{01000000}\textcolor{red}{00000000}00000000\textcolor{red}{00000000}00000000\textcolor{red}{00000000}00000000\textcolor{red}{00000000}00000000\textcolor{red}{00000000}00000000\newline\noindent
\textcolor{red}{00000000}00000000\textcolor{red}{00000000}00000000\textcolor{red}{00000000}\textbf{00100000}\textcolor{red}{00000000}00000000\textcolor{red}{00000000}00000000\textcolor{red}{00000000}00000000\textcolor{red}{00000000}00000000\textcolor{red}{00000000}00000000\newline\noindent
\textcolor{red}{00000000}00000000\textcolor{red}{00000000}00000000\textcolor{red}{00000000}\textbf{00010000}\textcolor{red}{00000000}00000000\textcolor{red}{00000000}00000000\textcolor{red}{00000000}00000000\textcolor{red}{00000000}00000000\textcolor{red}{00000000}00000000\newline\noindent
\textcolor{red}{00000000}00000000\textcolor{red}{00000000}00000000\textcolor{red}{00000000}\textbf{00001000}\textcolor{red}{00000000}00000000\textcolor{red}{00000000}00000000\textcolor{red}{00000000}00000000\textcolor{red}{00000000}00000000\textcolor{red}{00000000}00000000\newline\noindent
\textcolor{red}{00000000}00000000\textcolor{red}{00000000}00000000\textcolor{red}{00000000}\textbf{00000100}\textcolor{red}{00000000}00000000\textcolor{red}{00000000}00000000\textcolor{red}{00000000}00000000\textcolor{red}{00000000}00000000\textcolor{red}{00000000}00000000\newline\noindent
\textcolor{red}{00000000}00000000\textcolor{red}{00000000}00000000\textcolor{red}{00000000}\textbf{00000010}\textcolor{red}{00000000}00000000\textcolor{red}{00000000}00000000\textcolor{red}{00000000}00000000\textcolor{red}{00000000}00000000\textcolor{red}{00000000}00000000\newline\noindent
\textcolor{red}{00000000}00000000\textcolor{red}{00000000}00000000\textcolor{red}{00000000}\textbf{00000001}\textcolor{red}{00000000}00000000\textcolor{red}{00000000}00000000\textcolor{red}{00000000}00000000\textcolor{red}{00000000}00000000\textcolor{red}{00000000}00000000\newline\noindent
\end{ttfamily}
\end{center}

In the end, the equations of the function \texttt{ShiftRows} for the 128-bit of the block $B=(b_0 \dots b_{127})$ are:

\begin{small}
\begin{align*}
	(x_{0}, x_{1}, x_{2}, x_{3}, x_{4}, x_{5}, x_{6}, x_{7}, x_{40}, x_{41}, x_{42}, x_{43}, x_{44}, x_{45}, x_{46}, x_{47},\\
	x_{80}, x_{81}, x_{82}, x_{83}, x_{84}, x_{85}, x_{86}, x_{87}, x_{120}, x_{121}, x_{122}, x_{123}, x_{124}, x_{125}, x_{126}, x_{127},\\
	x_{32}, x_{33}, x_{34}, x_{35}, x_{36}, x_{37}, x_{38}, x_{39}, x_{72}, x_{73}, x_{74}, x_{75}, x_{76}, x_{77}, x_{78}, x_{79},\\
	x_{112}, x_{113}, x_{114}, x_{115}, x_{116}, x_{117}, x_{118}, x_{119}, x_{24}, x_{25}, x_{26}, x_{27}, x_{28}, x_{29}, x_{30}, x_{31},\\
	x_{64}, x_{65}, x_{66}, x_{67}, x_{68}, x_{69}, x_{70}, x_{71}, x_{104}, x_{105}, x_{106}, x_{107}, x_{108}, x_{109}, x_{110}, x_{111},\\
	x_{16}, x_{17}, x_{18}, x_{19}, x_{20}, x_{21}, x_{22}, x_{23}, x_{56}, x_{57}, x_{58}, x_{59}, x_{60}, x_{61}, x_{62}, x_{63},\\
	x_{96}, x_{97}, x_{98}, x_{99}, x_{100}, x_{101}, x_{102}, x_{103}, x_{8}, x_{9}, x_{10}, x_{11}, x_{12}, x_{13}, x_{14}, x_{15},\\
	x_{48}, x_{49}, x_{50}, x_{51}, x_{52}, x_{53}, x_{54}, x_{55}, x_{88}, x_{89}, x_{90}, x_{91}, x_{92}, x_{93}, x_{94}, x_{95})\\
\end{align*}
\end{small}

\subsubsection{Solution for MixColumns function}

The function \texttt{MixColumns} acts on the states array, column by column, treating each column as a polynomial with four terms. Each column is multiplied by a square matrix. For each column we have:

$$
\begin{pmatrix}
b_i' \\
b_{i+1}' \\
b_{i+2}' \\
b_{i+3}'
\end{pmatrix}
=
\begin{pmatrix}
02 & 03 & 01 & 01\\
01 & 02 & 03 & 01\\
01 & 01 & 02 & 03\\
03 & 01 & 01 & 02
\end{pmatrix}
\bullet
\begin{pmatrix}
b_i \\
b_{i+1} \\
b_{i+2} \\
b_{i+3}
\end{pmatrix}
$$

Thus, for the first byte of the column we have the equation:

$$b_i' = 02\bullet b_i\oplus 03\bullet b_{i+1}\oplus 01\bullet b_{i+2}\oplus 01\bullet b_{i+3}$$

As in $GF_2^8$, $01$ is the identity for multiplication, this equation becomes:

$$b_i' = 02\bullet b_i\oplus 03\bullet b_{i+1}\oplus b_{i+2}\oplus b_{i+3}$$

We have the same simplification for all equations describing the multiplication of the column of the states array by the square matrix. Therefore we only need to calculate truth tables for multiplication by $02$ and $03$ in $GF_2^8$.

For example, the equations of the bits $b_{120}$ to $b_{127}$ are the following:

\begin{align*}
	b_{120} &= x_{97} \oplus x_{96} \oplus x_{104} \oplus x_{112} \oplus x_{121} \\
	b_{121} &= x_{98} \oplus x_{97} \oplus x_{105} \oplus x_{113} \oplus x_{122} \\
	b_{122} &= x_{99} \oplus x_{98} \oplus x_{106} \oplus x_{114} \oplus x_{123} \\
	b_{123} &= x_{100} \oplus x_{99} \oplus x_{96} \oplus x_{107} \oplus x_{115} \oplus x_{124} \oplus x_{120} \\
	b_{124} &= x_{101} \oplus x_{100} \oplus x_{96} \oplus x_{108} \oplus x_{116} \oplus x_{125} \oplus x_{120} \\
	b_{125} &= x_{102} \oplus x_{101} \oplus x_{109} \oplus x_{117} \oplus x_{126} \\
	b_{126} &= x_{103} \oplus x_{102} \oplus x_{96} \oplus x_{110} \oplus x_{118} \oplus x_{127} \oplus x_{120} \\
	b_{127} &= x_{103} \oplus x_{96} \oplus x_{111} \oplus x_{119} \oplus x_{120} \\
\end{align*}

\subsubsection{Solution for the key expansion function}

To recall, in the algorithm of the AES-128, $\text{Nb}=4$ words and $\text{Nr}=10$ words, with 1 word = 4 bytes = 32 bits.

The function \texttt{AddRoundKey} adds a round key to the state table by a simple bitwise \texttt{XOR} operation. These rounds keys are computed by a key expansion function. This latter generates a set of $Nb (Nr + 1) = 44$ words of 32 bit that to say 11 keys of 128 bits derived from the first key. The algorithm used for the expansion of the key involves two functions \texttt{SubWord} and \texttt{RotWord} together with a round constant \texttt{Rcon}.

The generation of a global Boolean function for the key expansion algorithm is impossible because the generation of the key for the round $n$ involves the key of the round $n-1$. This interweaving of rounds keys does not allow us to generate a global Boolean function. On the other hand it is possible to generate a Boolean function corresponding to the calculation of a key of one round.

The first word $w_{i_0}$ of the round key $i$ is calculated according to the following equation: $$w_{i_0} = (SW \circ RW(w_{(i-1)_3})) \oplus Rcon_i \oplus w_{(i-1)_0}$$ with $SW()$ and $RW()$ respectively corresponding to the \texttt{SubWord} and \texttt{RotWord} functions.

The following words $w_{i_1}$, $w_{i_2}$ and $w_{i_3}$ are calculated according to the following equation: $$w_{i_n} = w_{i_{n-1}} \oplus w_{(i-1)_n}$$ with $1 \leq n \leq 3$.

The \texttt{SubWord} and \texttt{RotWord} functions are built on the same principle as the \texttt{SubBytes} and \texttt{ShiftRows} functions, thus we can reuse the methodology finalized previously.

In python language, the word generation function is written according to the following code \citelst{lst:keyword}.

\begin{lstlisting}[style=python, caption={Function for generating a key word in python},label=lst:keyword]
def generateWord(num):
	if (num < 4):
		w = generateGenericWord(wordSize*num, 'x')
	if (num >= 4):
		if ((num % 4) == 0):
			w = generateWord(3)
			w = rotWord(w)
			w = subWord(w, rconList[(num/4)-1])
			w = xorWords(w, generateWord(0))
		else:
			w = generateWord(num-1)
			w = xorWords(w, generateWord(num%4))
	return w
\end{lstlisting}

In this code, several scenarios are considered. The function \texttt{generateWord} takes in parameter the word number to generate, we know that this number is between 0 and 43. If the number is less than 4, the function returns the Boolean identity function as the first key used by the AES is the encryption key. If the number to modulo 4 is zero, the function returns a Boolean functions describing the composition of \texttt{SubWord} and \texttt{RotWord} functions and the application of the \texttt{XOR} with the \texttt{Rcon} constant. Finally, if the number to modulo 4 is not zero, the function returns the Boolean function describing the \texttt{XOR} with the corresponding word in the previous round.

We now have a Boolean function describing a round expansion of the key. As we have seen, the key expansion algorithm involves at round $n$ the keys of round $n-1$. To integrate our Boolean function in the encryption process of the AES, we must, at every round, add a temporary variable corresponding to the key of the previous round.

As an example, the Boolean equation of the bit $b_{0}$ of the fourth word on the 44 words generate by the key expansion process, is given in the figure \ref{fig:word} page~\pageref{fig:word}.

\begin{figure}
	\begin{flushleft}
	\footnotesize{%
	$x_{109} \oplus x_{109}x_{111} \oplus x_{109}x_{110} \oplus x_{108}x_{109}x_{111} \oplus x_{108}x_{109}x_{110} \oplus x_{108}x_{109}x_{110}x_{111} \oplus x_{107} \oplus x_{107}x_{110}x_{111} \oplus x_{107}x_{109} \oplus x_{107}x_{109}x_{110}x_{111} \oplus x_{107}x_{108}x_{110}x_{111} \oplus x_{107}x_{108}x_{109}x_{110} \oplus x_{107}x_{108}x_{109}x_{110}x_{111} \oplus x_{106} \oplus x_{106}x_{110}x_{111} \oplus x_{106}x_{109}x_{111} \oplus x_{106}x_{109}x_{110}x_{111} \oplus x_{106}x_{108} \oplus x_{106}x_{108}x_{111} \oplus x_{106}x_{108}x_{110} \oplus x_{106}x_{108}x_{109} \oplus x_{106}x_{108}x_{109}x_{111} \oplus x_{106}x_{108}x_{109}x_{110} \oplus x_{106}x_{107}x_{111} \oplus x_{106}x_{107}x_{109}x_{110} \oplus x_{106}x_{107}x_{108} \oplus x_{106}x_{107}x_{108}x_{110}x_{111} \oplus x_{106}x_{107}x_{108}x_{109}x_{111} \oplus x_{105}x_{111} \oplus x_{105}x_{110}x_{111} \oplus x_{105}x_{109} \oplus x_{105}x_{109}x_{110} \oplus x_{105}x_{108}x_{111} \oplus x_{105}x_{108}x_{110} \oplus x_{105}x_{108}x_{110}x_{111} \oplus x_{105}x_{108}x_{109}x_{111} \oplus x_{105}x_{108}x_{109}x_{110}x_{111} \oplus x_{105}x_{107} \oplus x_{105}x_{107}x_{109} \oplus x_{105}x_{107}x_{109}x_{111} \oplus x_{105}x_{107}x_{109}x_{110} \oplus x_{105}x_{107}x_{109}x_{110}x_{111} \oplus x_{105}x_{107}x_{108}x_{111} \oplus x_{105}x_{107}x_{108}x_{109}x_{111} \oplus x_{105}x_{106}x_{111} \oplus x_{105}x_{106}x_{109} \oplus x_{105}x_{106}x_{108}x_{111} \oplus x_{105}x_{106}x_{108}x_{109}x_{110} \oplus x_{105}x_{106}x_{107} \oplus x_{105}x_{106}x_{107}x_{110}x_{111} \oplus x_{105}x_{106}x_{107}x_{109}x_{110} \oplus x_{105}x_{106}x_{107}x_{108} \oplus x_{105}x_{106}x_{107}x_{108}x_{111} \oplus x_{105}x_{106}x_{107}x_{108}x_{109} \oplus x_{105}x_{106}x_{107}x_{108}x_{109}x_{111} \oplus x_{104} \oplus x_{104}x_{111} \oplus x_{104}x_{110} \oplus x_{104}x_{109}x_{111} \oplus x_{104}x_{109}x_{110}x_{111} \oplus x_{104}x_{108}x_{111} \oplus x_{104}x_{108}x_{109}x_{111} \oplus x_{104}x_{108}x_{109}x_{110} \oplus x_{104}x_{107}x_{110} \oplus x_{104}x_{107}x_{110}x_{111} \oplus x_{104}x_{107}x_{109}x_{111} \oplus x_{104}x_{107}x_{108}x_{111} \oplus x_{104}x_{107}x_{108}x_{110} \oplus x_{104}x_{107}x_{108}x_{110}x_{111} \oplus x_{104}x_{107}x_{108}x_{109} \oplus x_{104}x_{107}x_{108}x_{109}x_{111} \oplus x_{104}x_{106} \oplus x_{104}x_{106}x_{109}x_{110}x_{111} \oplus x_{104}x_{106}x_{108} \oplus x_{104}x_{106}x_{108}x_{111} \oplus x_{104}x_{106}x_{107} \oplus x_{104}x_{106}x_{107}x_{110} \oplus x_{104}x_{106}x_{107}x_{110}x_{111} \oplus x_{104}x_{106}x_{107}x_{109}x_{110}x_{111} \oplus x_{104}x_{106}x_{107}x_{108}x_{110}x_{111} \oplus x_{104}x_{106}x_{107}x_{108}x_{109}x_{111} \oplus x_{104}x_{105}x_{111} \oplus x_{104}x_{105}x_{109} \oplus x_{104}x_{105}x_{109}x_{110}x_{111} \oplus x_{104}x_{105}x_{108}x_{111} \oplus x_{104}x_{105}x_{108}x_{110} \oplus x_{104}x_{105}x_{108}x_{109}x_{110}x_{111} \oplus x_{104}x_{105}x_{107} \oplus x_{104}x_{105}x_{107}x_{111} \oplus x_{104}x_{105}x_{107}x_{110} \oplus x_{104}x_{105}x_{107}x_{109} \oplus x_{104}x_{105}x_{107}x_{109}x_{110} \oplus x_{104}x_{105}x_{107}x_{108}x_{111} \oplus x_{104}x_{105}x_{107}x_{108}x_{110}x_{111} \oplus x_{104}x_{105}x_{107}x_{108}x_{109}x_{111} \oplus x_{104}x_{105}x_{106}x_{110} \oplus x_{104}x_{105}x_{106}x_{110}x_{111} \oplus x_{104}x_{105}x_{106}x_{109} \oplus x_{104}x_{105}x_{106}x_{109}x_{110} \oplus x_{104}x_{105}x_{106}x_{108}x_{111} \oplus x_{104}x_{105}x_{106}x_{108}x_{110} \oplus x_{104}x_{105}x_{106}x_{108}x_{110}x_{111} \oplus x_{104}x_{105}x_{106}x_{108}x_{109}x_{111} \oplus x_{104}x_{105}x_{106}x_{107} \oplus x_{104}x_{105}x_{106}x_{107}x_{110} \oplus x_{104}x_{105}x_{106}x_{107}x_{109}x_{111} \oplus x_{104}x_{105}x_{106}x_{107}x_{108} \oplus x_{104}x_{105}x_{106}x_{107}x_{108}x_{110} \oplus x_{104}x_{105}x_{106}x_{107}x_{108}x_{110}x_{111} \oplus x_{104}x_{105}x_{106}x_{107}x_{108}x_{109}x_{111} \oplus x_{0}$
	}
	\end{flushleft}
	\caption{\'Equation of bit $b_{0}$ of the 4\textsuperscript{th} word}
	\label{fig:word}
\end{figure}

\subsubsection{Global solution}

We have now a Boolean function for each function \texttt{SubBytes} $SB()$, \texttt{ShiftRows} $SR()$ and \texttt{MixColumns} $MC()$. In the arrangement of one round, these functions are combined. So for a 128-bit block $B = (b_1, \cdots, b_{128})$ as output of the \texttt{AddRoundKey} function, the block $B' = (b'_1, \cdots, b'_{128})$ as output of the combination of these three functions is such that: $$B' = MC\circ SR\circ SB(B)$$

To realize the files as described above, it is necessary to reduce the composition of these three functions in one Boolean equation. To achieve this, we just have to replace each input variable of a function by the output value of the previous function using the following equation: $$b'_i = MC( SR( SB(b_i) ) ) \quad\forall i \in (1, \cdots, 128)$$

In python language, the round generation function is written according to the following code \citelst{lst:round}.

\begin{lstlisting}[style=python, caption={The equation for calculating an encryption round function},label=lst:round]
def writeRoundEnc(numRound, equaSB, equaSR, equaMC):
	printColor('## Round%s' % numRound, GREEN)
	resultSR = []
	resultMC = []
	for i in xrange(blockSize):
		equaSR[i] = equaSR[i].split('_')
		resultSR.append(equaSB[int(equaSR[i][1])])

	for i in xrange(blockSize):
		tmp = ''
		for monomial in equaMC[i].split('+'):
			tmp += resultSR[int(monomial.split('_')[1])]
			tmp += '+'
		resultMC.append(tmp.rstrip('+'))
	binMon = generateBinaryMonomes(resultMC)
	return resultMC
\end{lstlisting}

The Boolean equation of one round of the AES for the bit $b_{0}$ is given in the figure \ref{fig:round} page~\pageref{fig:round}.

\begin{figure}
	\begin{flushleft}
	\tiny{%
$1 \oplus x_{4} \oplus x_{4}x_{6} \oplus x_{4}x_{6}x_{7} \oplus x_{4}x_{5} \oplus x_{3}x_{7} \oplus x_{3}x_{6}x_{7} \oplus x_{3}x_{5}x_{7} \oplus x_{3}x_{5}x_{6}x_{7} \oplus x_{3}x_{4}x_{6} \oplus x_{3}x_{4}x_{6}x_{7} \oplus x_{3}x_{4}x_{5} \oplus x_{3}x_{4}x_{5}x_{6} \oplus x_{2} \oplus x_{2}x_{7} \oplus x_{2}x_{6}x_{7} \oplus x_{2}x_{5}x_{7} \oplus x_{2}x_{5}x_{6} \oplus x_{2}x_{5}x_{6}x_{7} \oplus x_{2}x_{4} \oplus x_{2}x_{4}x_{7} \oplus x_{2}x_{4}x_{5}x_{7} \oplus x_{2}x_{4}x_{5}x_{6} \oplus x_{2}x_{4}x_{5}x_{6}x_{7} \oplus x_{2}x_{3}x_{7} \oplus x_{2}x_{3}x_{5}x_{7} \oplus x_{2}x_{3}x_{5}x_{6} \oplus x_{2}x_{3}x_{4}x_{6}x_{7} \oplus x_{2}x_{3}x_{4}x_{5} \oplus x_{2}x_{3}x_{4}x_{5}x_{7} \oplus x_{2}x_{3}x_{4}x_{5}x_{6} \oplus x_{1} \oplus x_{1}x_{5}x_{7} \oplus x_{1}x_{5}x_{6} \oplus x_{1}x_{4}x_{7} \oplus x_{1}x_{4}x_{6} \oplus x_{1}x_{4}x_{6}x_{7} \oplus x_{1}x_{4}x_{5}x_{7} \oplus x_{1}x_{4}x_{5}x_{6} \oplus x_{1}x_{4}x_{5}x_{6}x_{7} \oplus x_{1}x_{3} \oplus x_{1}x_{3}x_{7} \oplus x_{1}x_{3}x_{6} \oplus x_{1}x_{3}x_{5} \oplus x_{1}x_{3}x_{5}x_{7} \oplus x_{1}x_{3}x_{4} \oplus x_{1}x_{3}x_{4}x_{7} \oplus x_{1}x_{3}x_{4}x_{6} \oplus x_{1}x_{3}x_{4}x_{5} \oplus x_{1}x_{2}x_{7} \oplus x_{1}x_{2}x_{6} \oplus x_{1}x_{2}x_{4}x_{5} \oplus x_{1}x_{2}x_{4}x_{5}x_{7} \oplus x_{1}x_{2}x_{4}x_{5}x_{6}x_{7} \oplus x_{1}x_{2}x_{3} \oplus x_{1}x_{2}x_{3}x_{5}x_{6} \oplus x_{1}x_{2}x_{3}x_{5}x_{6}x_{7} \oplus x_{1}x_{2}x_{3}x_{4}x_{6} \oplus x_{1}x_{2}x_{3}x_{4}x_{6}x_{7} \oplus x_{0}x_{7} \oplus x_{0}x_{6} \oplus x_{0}x_{5}x_{6} \oplus x_{0}x_{4} \oplus x_{0}x_{4}x_{6}x_{7} \oplus x_{0}x_{4}x_{5} \oplus x_{0}x_{4}x_{5}x_{7} \oplus x_{0}x_{4}x_{5}x_{6}x_{7} \oplus x_{0}x_{3}x_{7} \oplus x_{0}x_{3}x_{6} \oplus x_{0}x_{3}x_{5} \oplus x_{0}x_{3}x_{5}x_{6} \oplus x_{0}x_{3}x_{4}x_{7} \oplus x_{0}x_{3}x_{4}x_{6} \oplus x_{0}x_{3}x_{4}x_{5}x_{6} \oplus x_{0}x_{3}x_{4}x_{5}x_{6}x_{7} \oplus x_{0}x_{2} \oplus x_{0}x_{2}x_{7} \oplus x_{0}x_{2}x_{5}x_{7} \oplus x_{0}x_{2}x_{4} \oplus x_{0}x_{2}x_{4}x_{6} \oplus x_{0}x_{2}x_{4}x_{5} \oplus x_{0}x_{2}x_{4}x_{5}x_{6} \oplus x_{0}x_{2}x_{3}x_{6} \oplus x_{0}x_{2}x_{3}x_{6}x_{7} \oplus x_{0}x_{2}x_{3}x_{5}x_{7} \oplus x_{0}x_{2}x_{3}x_{5}x_{6}x_{7} \oplus x_{0}x_{2}x_{3}x_{4}x_{6} \oplus x_{0}x_{2}x_{3}x_{4}x_{6}x_{7} \oplus x_{0}x_{1}x_{6} \oplus x_{0}x_{1}x_{6}x_{7} \oplus x_{0}x_{1}x_{4} \oplus x_{0}x_{1}x_{4}x_{6}x_{7} \oplus x_{0}x_{1}x_{3}x_{7} \oplus x_{0}x_{1}x_{3}x_{6} \oplus x_{0}x_{1}x_{3}x_{5}x_{7} \oplus x_{0}x_{1}x_{3}x_{5}x_{6}x_{7} \oplus x_{0}x_{1}x_{3}x_{4}x_{7} \oplus x_{0}x_{1}x_{3}x_{4}x_{6}x_{7} \oplus x_{0}x_{1}x_{3}x_{4}x_{5} \oplus x_{0}x_{1}x_{2} \oplus x_{0}x_{1}x_{2}x_{5}x_{7} \oplus x_{0}x_{1}x_{2}x_{5}x_{6} \oplus x_{0}x_{1}x_{2}x_{5}x_{6}x_{7} \oplus x_{0}x_{1}x_{2}x_{4}x_{6}x_{7} \oplus x_{0}x_{1}x_{2}x_{4}x_{5} \oplus x_{0}x_{1}x_{2}x_{3} \oplus x_{0}x_{1}x_{2}x_{3}x_{7} \oplus x_{0}x_{1}x_{2}x_{3}x_{6} \oplus x_{0}x_{1}x_{2}x_{3}x_{5}x_{6}x_{7} \oplus x_{0}x_{1}x_{2}x_{3}x_{4} \oplus x_{0}x_{1}x_{2}x_{3}x_{4}x_{6} \oplus x_{0}x_{1}x_{2}x_{3}x_{4}x_{6}x_{7} \oplus 1 \oplus x_{44} \oplus x_{44}x_{46} \oplus x_{44}x_{46}x_{47} \oplus x_{44}x_{45} \oplus x_{43}x_{47} \oplus x_{43}x_{46}x_{47} \oplus x_{43}x_{45}x_{47} \oplus x_{43}x_{45}x_{46}x_{47} \oplus x_{43}x_{44}x_{46} \oplus x_{43}x_{44}x_{46}x_{47} \oplus x_{43}x_{44}x_{45} \oplus x_{43}x_{44}x_{45}x_{46} \oplus x_{42} \oplus x_{42}x_{47} \oplus x_{42}x_{46}x_{47} \oplus x_{42}x_{45}x_{47} \oplus x_{42}x_{45}x_{46} \oplus x_{42}x_{45}x_{46}x_{47} \oplus x_{42}x_{44} \oplus x_{42}x_{44}x_{47} \oplus x_{42}x_{44}x_{45}x_{47} \oplus x_{42}x_{44}x_{45}x_{46} \oplus x_{42}x_{44}x_{45}x_{46}x_{47} \oplus x_{42}x_{43}x_{47} \oplus x_{42}x_{43}x_{45}x_{47} \oplus x_{42}x_{43}x_{45}x_{46} \oplus x_{42}x_{43}x_{44}x_{46}x_{47} \oplus x_{42}x_{43}x_{44}x_{45} \oplus x_{42}x_{43}x_{44}x_{45}x_{47} \oplus x_{42}x_{43}x_{44}x_{45}x_{46} \oplus x_{41} \oplus x_{41}x_{45}x_{47} \oplus x_{41}x_{45}x_{46} \oplus x_{41}x_{44}x_{47} \oplus x_{41}x_{44}x_{46} \oplus x_{41}x_{44}x_{46}x_{47} \oplus x_{41}x_{44}x_{45}x_{47} \oplus x_{41}x_{44}x_{45}x_{46} \oplus x_{41}x_{44}x_{45}x_{46}x_{47} \oplus x_{41}x_{43} \oplus x_{41}x_{43}x_{47} \oplus x_{41}x_{43}x_{46} \oplus x_{41}x_{43}x_{45} \oplus x_{41}x_{43}x_{45}x_{47} \oplus x_{41}x_{43}x_{44} \oplus x_{41}x_{43}x_{44}x_{47} \oplus x_{41}x_{43}x_{44}x_{46} \oplus x_{41}x_{43}x_{44}x_{45} \oplus x_{41}x_{42}x_{47} \oplus x_{41}x_{42}x_{46} \oplus x_{41}x_{42}x_{44}x_{45} \oplus x_{41}x_{42}x_{44}x_{45}x_{47} \oplus x_{41}x_{42}x_{44}x_{45}x_{46}x_{47} \oplus x_{41}x_{42}x_{43} \oplus x_{41}x_{42}x_{43}x_{45}x_{46} \oplus x_{41}x_{42}x_{43}x_{45}x_{46}x_{47} \oplus x_{41}x_{42}x_{43}x_{44}x_{46} \oplus x_{41}x_{42}x_{43}x_{44}x_{46}x_{47} \oplus x_{40}x_{47} \oplus x_{40}x_{46} \oplus x_{40}x_{45}x_{46} \oplus x_{40}x_{44} \oplus x_{40}x_{44}x_{46}x_{47} \oplus x_{40}x_{44}x_{45} \oplus x_{40}x_{44}x_{45}x_{47} \oplus x_{40}x_{44}x_{45}x_{46}x_{47} \oplus x_{40}x_{43}x_{47} \oplus x_{40}x_{43}x_{46} \oplus x_{40}x_{43}x_{45} \oplus x_{40}x_{43}x_{45}x_{46} \oplus x_{40}x_{43}x_{44}x_{47} \oplus x_{40}x_{43}x_{44}x_{46} \oplus x_{40}x_{43}x_{44}x_{45}x_{46} \oplus x_{40}x_{43}x_{44}x_{45}x_{46}x_{47} \oplus x_{40}x_{42} \oplus x_{40}x_{42}x_{47} \oplus x_{40}x_{42}x_{45}x_{47} \oplus x_{40}x_{42}x_{44} \oplus x_{40}x_{42}x_{44}x_{46} \oplus x_{40}x_{42}x_{44}x_{45} \oplus x_{40}x_{42}x_{44}x_{45}x_{46} \oplus x_{40}x_{42}x_{43}x_{46} \oplus x_{40}x_{42}x_{43}x_{46}x_{47} \oplus x_{40}x_{42}x_{43}x_{45}x_{47} \oplus x_{40}x_{42}x_{43}x_{45}x_{46}x_{47} \oplus x_{40}x_{42}x_{43}x_{44}x_{46} \oplus x_{40}x_{42}x_{43}x_{44}x_{46}x_{47} \oplus x_{40}x_{41}x_{46} \oplus x_{40}x_{41}x_{46}x_{47} \oplus x_{40}x_{41}x_{44} \oplus x_{40}x_{41}x_{44}x_{46}x_{47} \oplus x_{40}x_{41}x_{43}x_{47} \oplus x_{40}x_{41}x_{43}x_{46} \oplus x_{40}x_{41}x_{43}x_{45}x_{47} \oplus x_{40}x_{41}x_{43}x_{45}x_{46}x_{47} \oplus x_{40}x_{41}x_{43}x_{44}x_{47} \oplus x_{40}x_{41}x_{43}x_{44}x_{46}x_{47} \oplus x_{40}x_{41}x_{43}x_{44}x_{45} \oplus x_{40}x_{41}x_{42} \oplus x_{40}x_{41}x_{42}x_{45}x_{47} \oplus x_{40}x_{41}x_{42}x_{45}x_{46} \oplus x_{40}x_{41}x_{42}x_{45}x_{46}x_{47} \oplus x_{40}x_{41}x_{42}x_{44}x_{46}x_{47} \oplus x_{40}x_{41}x_{42}x_{44}x_{45} \oplus x_{40}x_{41}x_{42}x_{43} \oplus x_{40}x_{41}x_{42}x_{43}x_{47} \oplus x_{40}x_{41}x_{42}x_{43}x_{46} \oplus x_{40}x_{41}x_{42}x_{43}x_{45}x_{46}x_{47} \oplus x_{40}x_{41}x_{42}x_{43}x_{44} \oplus x_{40}x_{41}x_{42}x_{43}x_{44}x_{46} \oplus x_{40}x_{41}x_{42}x_{43}x_{44}x_{46}x_{47} \oplus x_{45} \oplus x_{45}x_{47} \oplus x_{45}x_{46} \oplus x_{44}x_{45}x_{47} \oplus x_{44}x_{45}x_{46} \oplus x_{44}x_{45}x_{46}x_{47} \oplus x_{43} \oplus x_{43}x_{46}x_{47} \oplus x_{43}x_{45} \oplus x_{43}x_{45}x_{46}x_{47} \oplus x_{43}x_{44}x_{46}x_{47} \oplus x_{43}x_{44}x_{45}x_{46} \oplus x_{43}x_{44}x_{45}x_{46}x_{47} \oplus x_{42} \oplus x_{42}x_{46}x_{47} \oplus x_{42}x_{45}x_{47} \oplus x_{42}x_{45}x_{46}x_{47} \oplus x_{42}x_{44} \oplus x_{42}x_{44}x_{47} \oplus x_{42}x_{44}x_{46} \oplus x_{42}x_{44}x_{45} \oplus x_{42}x_{44}x_{45}x_{47} \oplus x_{42}x_{44}x_{45}x_{46} \oplus x_{42}x_{43}x_{47} \oplus x_{42}x_{43}x_{45}x_{46} \oplus x_{42}x_{43}x_{44} \oplus x_{42}x_{43}x_{44}x_{46}x_{47} \oplus x_{42}x_{43}x_{44}x_{45}x_{47} \oplus x_{41}x_{47} \oplus x_{41}x_{46}x_{47} \oplus x_{41}x_{45} \oplus x_{41}x_{45}x_{46} \oplus x_{41}x_{44}x_{47} \oplus x_{41}x_{44}x_{46} \oplus x_{41}x_{44}x_{46}x_{47} \oplus x_{41}x_{44}x_{45}x_{47} \oplus x_{41}x_{44}x_{45}x_{46}x_{47} \oplus x_{41}x_{43} \oplus x_{41}x_{43}x_{45} \oplus x_{41}x_{43}x_{45}x_{47} \oplus x_{41}x_{43}x_{45}x_{46} \oplus x_{41}x_{43}x_{45}x_{46}x_{47} \oplus x_{41}x_{43}x_{44}x_{47} \oplus x_{41}x_{43}x_{44}x_{45}x_{47} \oplus x_{41}x_{42}x_{47} \oplus x_{41}x_{42}x_{45} \oplus x_{41}x_{42}x_{44}x_{47} \oplus x_{41}x_{42}x_{44}x_{45}x_{46} \oplus x_{41}x_{42}x_{43} \oplus x_{41}x_{42}x_{43}x_{46}x_{47} \oplus x_{41}x_{42}x_{43}x_{45}x_{46} \oplus x_{41}x_{42}x_{43}x_{44} \oplus x_{41}x_{42}x_{43}x_{44}x_{47} \oplus x_{41}x_{42}x_{43}x_{44}x_{45} \oplus x_{41}x_{42}x_{43}x_{44}x_{45}x_{47} \oplus x_{40} \oplus x_{40}x_{47} \oplus x_{40}x_{46} \oplus x_{40}x_{45}x_{47} \oplus x_{40}x_{45}x_{46}x_{47} \oplus x_{40}x_{44}x_{47} \oplus x_{40}x_{44}x_{45}x_{47} \oplus x_{40}x_{44}x_{45}x_{46} \oplus x_{40}x_{43}x_{46} \oplus x_{40}x_{43}x_{46}x_{47} \oplus x_{40}x_{43}x_{45}x_{47} \oplus x_{40}x_{43}x_{44}x_{47} \oplus x_{40}x_{43}x_{44}x_{46} \oplus x_{40}x_{43}x_{44}x_{46}x_{47} \oplus x_{40}x_{43}x_{44}x_{45} \oplus x_{40}x_{43}x_{44}x_{45}x_{47} \oplus x_{40}x_{42} \oplus x_{40}x_{42}x_{45}x_{46}x_{47} \oplus x_{40}x_{42}x_{44} \oplus x_{40}x_{42}x_{44}x_{47} \oplus x_{40}x_{42}x_{43} \oplus x_{40}x_{42}x_{43}x_{46} \oplus x_{40}x_{42}x_{43}x_{46}x_{47} \oplus x_{40}x_{42}x_{43}x_{45}x_{46}x_{47} \oplus x_{40}x_{42}x_{43}x_{44}x_{46}x_{47} \oplus x_{40}x_{42}x_{43}x_{44}x_{45}x_{47} \oplus x_{40}x_{41}x_{47} \oplus x_{40}x_{41}x_{45} \oplus x_{40}x_{41}x_{45}x_{46}x_{47} \oplus x_{40}x_{41}x_{44}x_{47} \oplus x_{40}x_{41}x_{44}x_{46} \oplus x_{40}x_{41}x_{44}x_{45}x_{46}x_{47} \oplus x_{40}x_{41}x_{43} \oplus x_{40}x_{41}x_{43}x_{47} \oplus x_{40}x_{41}x_{43}x_{46} \oplus x_{40}x_{41}x_{43}x_{45} \oplus x_{40}x_{41}x_{43}x_{45}x_{46} \oplus x_{40}x_{41}x_{43}x_{44}x_{47} \oplus x_{40}x_{41}x_{43}x_{44}x_{46}x_{47} \oplus x_{40}x_{41}x_{43}x_{44}x_{45}x_{47} \oplus x_{40}x_{41}x_{42}x_{46} \oplus x_{40}x_{41}x_{42}x_{46}x_{47} \oplus x_{40}x_{41}x_{42}x_{45} \oplus x_{40}x_{41}x_{42}x_{45}x_{46} \oplus x_{40}x_{41}x_{42}x_{44}x_{47} \oplus x_{40}x_{41}x_{42}x_{44}x_{46} \oplus x_{40}x_{41}x_{42}x_{44}x_{46}x_{47} \oplus x_{40}x_{41}x_{42}x_{44}x_{45}x_{47} \oplus x_{40}x_{41}x_{42}x_{43} \oplus x_{40}x_{41}x_{42}x_{43}x_{46} \oplus x_{40}x_{41}x_{42}x_{43}x_{45}x_{47} \oplus x_{40}x_{41}x_{42}x_{43}x_{44} \oplus x_{40}x_{41}x_{42}x_{43}x_{44}x_{46} \oplus x_{40}x_{41}x_{42}x_{43}x_{44}x_{46}x_{47} \oplus x_{40}x_{41}x_{42}x_{43}x_{44}x_{45}x_{47} \oplus x_{85} \oplus x_{85}x_{87} \oplus x_{85}x_{86} \oplus x_{84}x_{85}x_{87} \oplus x_{84}x_{85}x_{86} \oplus x_{84}x_{85}x_{86}x_{87} \oplus x_{83} \oplus x_{83}x_{86}x_{87} \oplus x_{83}x_{85} \oplus x_{83}x_{85}x_{86}x_{87} \oplus x_{83}x_{84}x_{86}x_{87} \oplus x_{83}x_{84}x_{85}x_{86} \oplus x_{83}x_{84}x_{85}x_{86}x_{87} \oplus x_{82} \oplus x_{82}x_{86}x_{87} \oplus x_{82}x_{85}x_{87} \oplus x_{82}x_{85}x_{86}x_{87} \oplus x_{82}x_{84} \oplus x_{82}x_{84}x_{87} \oplus x_{82}x_{84}x_{86} \oplus x_{82}x_{84}x_{85} \oplus x_{82}x_{84}x_{85}x_{87} \oplus x_{82}x_{84}x_{85}x_{86} \oplus x_{82}x_{83}x_{87} \oplus x_{82}x_{83}x_{85}x_{86} \oplus x_{82}x_{83}x_{84} \oplus x_{82}x_{83}x_{84}x_{86}x_{87} \oplus x_{82}x_{83}x_{84}x_{85}x_{87} \oplus x_{81}x_{87} \oplus x_{81}x_{86}x_{87} \oplus x_{81}x_{85} \oplus x_{81}x_{85}x_{86} \oplus x_{81}x_{84}x_{87} \oplus x_{81}x_{84}x_{86} \oplus x_{81}x_{84}x_{86}x_{87} \oplus x_{81}x_{84}x_{85}x_{87} \oplus x_{81}x_{84}x_{85}x_{86}x_{87} \oplus x_{81}x_{83} \oplus x_{81}x_{83}x_{85} \oplus x_{81}x_{83}x_{85}x_{87} \oplus x_{81}x_{83}x_{85}x_{86} \oplus x_{81}x_{83}x_{85}x_{86}x_{87} \oplus x_{81}x_{83}x_{84}x_{87} \oplus x_{81}x_{83}x_{84}x_{85}x_{87} \oplus x_{81}x_{82}x_{87} \oplus x_{81}x_{82}x_{85} \oplus x_{81}x_{82}x_{84}x_{87} \oplus x_{81}x_{82}x_{84}x_{85}x_{86} \oplus x_{81}x_{82}x_{83} \oplus x_{81}x_{82}x_{83}x_{86}x_{87} \oplus x_{81}x_{82}x_{83}x_{85}x_{86} \oplus x_{81}x_{82}x_{83}x_{84} \oplus x_{81}x_{82}x_{83}x_{84}x_{87} \oplus x_{81}x_{82}x_{83}x_{84}x_{85} \oplus x_{81}x_{82}x_{83}x_{84}x_{85}x_{87} \oplus x_{80} \oplus x_{80}x_{87} \oplus x_{80}x_{86} \oplus x_{80}x_{85}x_{87} \oplus x_{80}x_{85}x_{86}x_{87} \oplus x_{80}x_{84}x_{87} \oplus x_{80}x_{84}x_{85}x_{87} \oplus x_{80}x_{84}x_{85}x_{86} \oplus x_{80}x_{83}x_{86} \oplus x_{80}x_{83}x_{86}x_{87} \oplus x_{80}x_{83}x_{85}x_{87} \oplus x_{80}x_{83}x_{84}x_{87} \oplus x_{80}x_{83}x_{84}x_{86} \oplus x_{80}x_{83}x_{84}x_{86}x_{87} \oplus x_{80}x_{83}x_{84}x_{85} \oplus x_{80}x_{83}x_{84}x_{85}x_{87} \oplus x_{80}x_{82} \oplus x_{80}x_{82}x_{85}x_{86}x_{87} \oplus x_{80}x_{82}x_{84} \oplus x_{80}x_{82}x_{84}x_{87} \oplus x_{80}x_{82}x_{83} \oplus x_{80}x_{82}x_{83}x_{86} \oplus x_{80}x_{82}x_{83}x_{86}x_{87} \oplus x_{80}x_{82}x_{83}x_{85}x_{86}x_{87} \oplus x_{80}x_{82}x_{83}x_{84}x_{86}x_{87} \oplus x_{80}x_{82}x_{83}x_{84}x_{85}x_{87} \oplus x_{80}x_{81}x_{87} \oplus x_{80}x_{81}x_{85} \oplus x_{80}x_{81}x_{85}x_{86}x_{87} \oplus x_{80}x_{81}x_{84}x_{87} \oplus x_{80}x_{81}x_{84}x_{86} \oplus x_{80}x_{81}x_{84}x_{85}x_{86}x_{87} \oplus x_{80}x_{81}x_{83} \oplus x_{80}x_{81}x_{83}x_{87} \oplus x_{80}x_{81}x_{83}x_{86} \oplus x_{80}x_{81}x_{83}x_{85} \oplus x_{80}x_{81}x_{83}x_{85}x_{86} \oplus x_{80}x_{81}x_{83}x_{84}x_{87} \oplus x_{80}x_{81}x_{83}x_{84}x_{86}x_{87} \oplus x_{80}x_{81}x_{83}x_{84}x_{85}x_{87} \oplus x_{80}x_{81}x_{82}x_{86} \oplus x_{80}x_{81}x_{82}x_{86}x_{87} \oplus x_{80}x_{81}x_{82}x_{85} \oplus x_{80}x_{81}x_{82}x_{85}x_{86} \oplus x_{80}x_{81}x_{82}x_{84}x_{87} \oplus x_{80}x_{81}x_{82}x_{84}x_{86} \oplus x_{80}x_{81}x_{82}x_{84}x_{86}x_{87} \oplus x_{80}x_{81}x_{82}x_{84}x_{85}x_{87} \oplus x_{80}x_{81}x_{82}x_{83} \oplus x_{80}x_{81}x_{82}x_{83}x_{86} \oplus x_{80}x_{81}x_{82}x_{83}x_{85}x_{87} \oplus x_{80}x_{81}x_{82}x_{83}x_{84} \oplus x_{80}x_{81}x_{82}x_{83}x_{84}x_{86} \oplus x_{80}x_{81}x_{82}x_{83}x_{84}x_{86}x_{87} \oplus x_{80}x_{81}x_{82}x_{83}x_{84}x_{85}x_{87} \oplus x_{125} \oplus x_{125}x_{127} \oplus x_{125}x_{126} \oplus x_{124}x_{125}x_{127} \oplus x_{124}x_{125}x_{126} \oplus x_{124}x_{125}x_{126}x_{127} \oplus x_{123} \oplus x_{123}x_{126}x_{127} \oplus x_{123}x_{125} \oplus x_{123}x_{125}x_{126}x_{127} \oplus x_{123}x_{124}x_{126}x_{127} \oplus x_{123}x_{124}x_{125}x_{126} \oplus x_{123}x_{124}x_{125}x_{126}x_{127} \oplus x_{122} \oplus x_{122}x_{126}x_{127} \oplus x_{122}x_{125}x_{127} \oplus x_{122}x_{125}x_{126}x_{127} \oplus x_{122}x_{124} \oplus x_{122}x_{124}x_{127} \oplus x_{122}x_{124}x_{126} \oplus x_{122}x_{124}x_{125} \oplus x_{122}x_{124}x_{125}x_{127} \oplus x_{122}x_{124}x_{125}x_{126} \oplus x_{122}x_{123}x_{127} \oplus x_{122}x_{123}x_{125}x_{126} \oplus x_{122}x_{123}x_{124} \oplus x_{122}x_{123}x_{124}x_{126}x_{127} \oplus x_{122}x_{123}x_{124}x_{125}x_{127} \oplus x_{121}x_{127} \oplus x_{121}x_{126}x_{127} \oplus x_{121}x_{125} \oplus x_{121}x_{125}x_{126} \oplus x_{121}x_{124}x_{127} \oplus x_{121}x_{124}x_{126} \oplus x_{121}x_{124}x_{126}x_{127} \oplus x_{121}x_{124}x_{125}x_{127} \oplus x_{121}x_{124}x_{125}x_{126}x_{127} \oplus x_{121}x_{123} \oplus x_{121}x_{123}x_{125} \oplus x_{121}x_{123}x_{125}x_{127} \oplus x_{121}x_{123}x_{125}x_{126} \oplus x_{121}x_{123}x_{125}x_{126}x_{127} \oplus x_{121}x_{123}x_{124}x_{127} \oplus x_{121}x_{123}x_{124}x_{125}x_{127} \oplus x_{121}x_{122}x_{127} \oplus x_{121}x_{122}x_{125} \oplus x_{121}x_{122}x_{124}x_{127} \oplus x_{121}x_{122}x_{124}x_{125}x_{126} \oplus x_{121}x_{122}x_{123} \oplus x_{121}x_{122}x_{123}x_{126}x_{127} \oplus x_{121}x_{122}x_{123}x_{125}x_{126} \oplus x_{121}x_{122}x_{123}x_{124} \oplus x_{121}x_{122}x_{123}x_{124}x_{127} \oplus x_{121}x_{122}x_{123}x_{124}x_{125} \oplus x_{121}x_{122}x_{123}x_{124}x_{125}x_{127} \oplus x_{120} \oplus x_{120}x_{127} \oplus x_{120}x_{126} \oplus x_{120}x_{125}x_{127} \oplus x_{120}x_{125}x_{126}x_{127} \oplus x_{120}x_{124}x_{127} \oplus x_{120}x_{124}x_{125}x_{127} \oplus x_{120}x_{124}x_{125}x_{126} \oplus x_{120}x_{123}x_{126} \oplus x_{120}x_{123}x_{126}x_{127} \oplus x_{120}x_{123}x_{125}x_{127} \oplus x_{120}x_{123}x_{124}x_{127} \oplus x_{120}x_{123}x_{124}x_{126} \oplus x_{120}x_{123}x_{124}x_{126}x_{127} \oplus x_{120}x_{123}x_{124}x_{125} \oplus x_{120}x_{123}x_{124}x_{125}x_{127} \oplus x_{120}x_{122} \oplus x_{120}x_{122}x_{125}x_{126}x_{127} \oplus x_{120}x_{122}x_{124} \oplus x_{120}x_{122}x_{124}x_{127} \oplus x_{120}x_{122}x_{123} \oplus x_{120}x_{122}x_{123}x_{126} \oplus x_{120}x_{122}x_{123}x_{126}x_{127} \oplus x_{120}x_{122}x_{123}x_{125}x_{126}x_{127} \oplus x_{120}x_{122}x_{123}x_{124}x_{126}x_{127} \oplus x_{120}x_{122}x_{123}x_{124}x_{125}x_{127} \oplus x_{120}x_{121}x_{127} \oplus x_{120}x_{121}x_{125} \oplus x_{120}x_{121}x_{125}x_{126}x_{127} \oplus x_{120}x_{121}x_{124}x_{127} \oplus x_{120}x_{121}x_{124}x_{126} \oplus x_{120}x_{121}x_{124}x_{125}x_{126}x_{127} \oplus x_{120}x_{121}x_{123} \oplus x_{120}x_{121}x_{123}x_{127} \oplus x_{120}x_{121}x_{123}x_{126} \oplus x_{120}x_{121}x_{123}x_{125} \oplus x_{120}x_{121}x_{123}x_{125}x_{126} \oplus x_{120}x_{121}x_{123}x_{124}x_{127} \oplus x_{120}x_{121}x_{123}x_{124}x_{126}x_{127} \oplus x_{120}x_{121}x_{123}x_{124}x_{125}x_{127} \oplus x_{120}x_{121}x_{122}x_{126} \oplus x_{120}x_{121}x_{122}x_{126}x_{127} \oplus x_{120}x_{121}x_{122}x_{125} \oplus x_{120}x_{121}x_{122}x_{125}x_{126} \oplus x_{120}x_{121}x_{122}x_{124}x_{127} \oplus x_{120}x_{121}x_{122}x_{124}x_{126} \oplus x_{120}x_{121}x_{122}x_{124}x_{126}x_{127} \oplus x_{120}x_{121}x_{122}x_{124}x_{125}x_{127} \oplus x_{120}x_{121}x_{122}x_{123} \oplus x_{120}x_{121}x_{122}x_{123}x_{126} \oplus x_{120}x_{121}x_{122}x_{123}x_{125}x_{127} \oplus x_{120}x_{121}x_{122}x_{123}x_{124} \oplus x_{120}x_{121}x_{122}x_{123}x_{124}x_{126} \oplus x_{120}x_{121}x_{122}x_{123}x_{124}x_{126}x_{127} \oplus x_{120}x_{121}x_{122}x_{123}x_{124}x_{125}x_{127}$
	}
	\end{flushleft}
	\caption{Equation of the bit $b_{0}$ for one round}
	\label{fig:round}
\end{figure}

Finally, we can now describe under the form of Boolean equations the full process of AES encryption. The function in python language computing this process is given in Listing \ref{lst:ciphering} page~\pageref{lst:ciphering}.

\begin{lstlisting}[style=python, caption={Calculation of Boolean functions of the AES encryption process},label=lst:ciphering]
def generateEncFullFiles():
	printColor('## Ciphering process', YELLOW)
	createAESFiles('enc')
	addRoundKey(0, 'enc')
	writeRoundEnc(0, subBytes(), shiftRows(), mixColumns())
	addRoundKey(1, 'enc')
	writeRoundEnc(1, subBytes(), shiftRows(), mixColumns())
	addRoundKey(2, 'enc')
	writeRoundEnc(2, subBytes(), shiftRows(), mixColumns())
	addRoundKey(3, 'enc')
	writeRoundEnc(3, subBytes(), shiftRows(), mixColumns())
	addRoundKey(4, 'enc')
	writeRoundEnc(4, subBytes(), shiftRows(), mixColumns())
	addRoundKey(5, 'enc')
	writeRoundEnc(5, subBytes(), shiftRows(), mixColumns())
	addRoundKey(6, 'enc')
	writeRoundEnc(6, subBytes(), shiftRows(), mixColumns())
	addRoundKey(7, 'enc')
	writeRoundEnc(7, subBytes(), shiftRows(), mixColumns())
	addRoundKey(8, 'enc')
	writeRoundEnc(8, subBytes(), shiftRows(), mixColumns())
	addRoundKey(9, 'enc')
	writeFinalRoundEnc(9, subBytes(), shiftRows())
	addRoundKey(10, 'enc')
	writeEndFlag('enc')
	printColor('## Files generated', YELLOW)
\end{lstlisting}

\subsection{The equations for deciphering functions}

We will now detail the solution implemented for each of the sub-functions of the AES decryption algorithm.

\subsubsection{Solution for the round function}

The AES deciphering algorithm uses the \texttt{Inv\-ShiftRows}, \texttt{Inv\-SubBytes} and \texttt{Inv\-MixColumns} functions. Those functions are respectively the inverse functions of \texttt{ShiftRows}, \texttt{SubBytes} and \texttt{MixColumns} functions, used in the ciphering process. The pseudo code of the decryption function can be written as follows \citefig{fig:decipherpseudocode}, \texttt{Nb} corresponding to the 32-bits words numbe and \texttt{Nr} corresponding to the rounds number used in the algorithm.

\begin{figure}
	\begin{center}
	\begin{algorithmic}[1]
		\Function{InvCipher}{byte in[4*Nb], byte out[4*Nb], word w[Nb*(Nr+1)]}
			\State byte state[4,Nb]
			\State state $\gets$ in
			\State AddRounkey(state, w[Nr*Nb, (Nr+1)*Nb-1])
			\For {round=Nr-1 step -1 downto 1}
				\State InvShiftRows(state)
				\State InvSubBytes(state)
				\State AddRoundKey(state, w[round*Nb, (round+1)*Nb-1])
				\State InvMixColumns(state)
			\EndFor
			\State InvShiftRows(state)
			\State InvSubBytes(state)
			\State AddRounkey(state, w[0, Nb-1])
			\State \Return state
		\EndFunction
	\end{algorithmic}
	\end{center}
	\caption{Deciphering pseudo code}
	\label{fig:decipherpseudocode}
\end{figure}

The internal mechanisms to the three functions used in the round during decryption are similar to encryption functions. So we use the same reasoning as the one implemented earlier to generate the corresponding Boolean equations.

For example, the Boolean equation of the three transformations used in the deciphering process for the bit $b_{0}$ are given in figure \ref{fig:threefunctions} page~\pageref{fig:threefunctions}.

\begin{figure}
	\begin{flushleft}
	{\footnotesize{%
	$invSubBytes(b_0) = x_{6}x_{7} \oplus x_{5}x_{6} \oplus x_{4} \oplus x_{4}x_{7} \oplus x_{4}x_{5}x_{7} \oplus x_{4}x_{5}x_{6} \oplus x_{4}x_{5}x_{6}x_{7} \oplus x_{3}x_{7} \oplus x_{3}x_{6}x_{7} \oplus x_{3}x_{5} \oplus x_{3}x_{5}x_{6} \oplus x_{3}x_{5}x_{6}x_{7} \oplus x_{3}x_{4} \oplus x_{3}x_{4}x_{7} \oplus x_{3}x_{4}x_{6}x_{7} \oplus x_{3}x_{4}x_{5}x_{6} \oplus x_{3}x_{4}x_{5}x_{6}x_{7} \oplus x_{2}x_{6} \oplus x_{2}x_{5} \oplus x_{2}x_{5}x_{6} \oplus x_{2}x_{5}x_{6}x_{7} \oplus x_{2}x_{4}x_{6} \oplus x_{2}x_{4}x_{6}x_{7} \oplus x_{2}x_{4}x_{5}x_{7} \oplus x_{2}x_{3}x_{7} \oplus x_{2}x_{3}x_{6}x_{7} \oplus x_{2}x_{3}x_{5}x_{7} \oplus x_{2}x_{3}x_{5}x_{6}x_{7} \oplus x_{2}x_{3}x_{4}x_{6} \oplus x_{2}x_{3}x_{4}x_{5}x_{6} \oplus x_{1}x_{7} \oplus x_{1}x_{6} \oplus x_{1}x_{6}x_{7} \oplus x_{1}x_{5} \oplus x_{1}x_{4}x_{6}x_{7} \oplus x_{1}x_{4}x_{5}x_{7} \oplus x_{1}x_{3}x_{6} \oplus x_{1}x_{3}x_{6}x_{7} \oplus x_{1}x_{3}x_{5} \oplus x_{1}x_{3}x_{5}x_{6}x_{7} \oplus x_{1}x_{2} \oplus x_{1}x_{2}x_{7} \oplus x_{1}x_{2}x_{6}x_{7} \oplus x_{1}x_{2}x_{5}x_{6}x_{7} \oplus x_{1}x_{2}x_{4} \oplus x_{1}x_{2}x_{4}x_{7} \oplus x_{1}x_{2}x_{4}x_{6}x_{7} \oplus x_{1}x_{2}x_{4}x_{5}x_{7} \oplus x_{1}x_{2}x_{4}x_{5}x_{6} \oplus x_{1}x_{2}x_{4}x_{5}x_{6}x_{7} \oplus x_{1}x_{2}x_{3}x_{7} \oplus x_{1}x_{2}x_{3}x_{5}x_{7} \oplus x_{1}x_{2}x_{3}x_{5}x_{6} \oplus x_{1}x_{2}x_{3}x_{4} \oplus x_{1}x_{2}x_{3}x_{4}x_{6} \oplus x_{1}x_{2}x_{3}x_{4}x_{6}x_{7} \oplus x_{1}x_{2}x_{3}x_{4}x_{5} \oplus x_{1}x_{2}x_{3}x_{4}x_{5}x_{6} \oplus x_{0}x_{7} \oplus x_{0}x_{5}x_{7} \oplus x_{0}x_{5}x_{6}x_{7} \oplus x_{0}x_{4}x_{6}x_{7} \oplus x_{0}x_{4}x_{5}x_{6}x_{7} \oplus x_{0}x_{3} \oplus x_{0}x_{3}x_{6} \oplus x_{0}x_{3}x_{5} \oplus x_{0}x_{3}x_{5}x_{7} \oplus x_{0}x_{3}x_{5}x_{6}x_{7} \oplus x_{0}x_{3}x_{4}x_{6}x_{7} \oplus x_{0}x_{3}x_{4}x_{5} \oplus x_{0}x_{3}x_{4}x_{5}x_{7} \oplus x_{0}x_{2}x_{6} \oplus x_{0}x_{2}x_{5} \oplus x_{0}x_{2}x_{5}x_{6} \oplus x_{0}x_{2}x_{5}x_{6}x_{7} \oplus x_{0}x_{2}x_{4} \oplus x_{0}x_{2}x_{4}x_{7} \oplus x_{0}x_{2}x_{4}x_{6} \oplus x_{0}x_{2}x_{4}x_{5} \oplus x_{0}x_{2}x_{4}x_{5}x_{7} \oplus x_{0}x_{2}x_{3}x_{5}x_{6}x_{7} \oplus x_{0}x_{2}x_{3}x_{4} \oplus x_{0}x_{2}x_{3}x_{4}x_{6}x_{7} \oplus x_{0}x_{2}x_{3}x_{4}x_{5} \oplus x_{0}x_{2}x_{3}x_{4}x_{5}x_{6} \oplus x_{0}x_{1}x_{7} \oplus x_{0}x_{1}x_{5}x_{7} \oplus x_{0}x_{1}x_{5}x_{6}x_{7} \oplus x_{0}x_{1}x_{4}x_{7} \oplus x_{0}x_{1}x_{4}x_{6}x_{7} \oplus x_{0}x_{1}x_{4}x_{5}x_{6}x_{7} \oplus x_{0}x_{1}x_{3} \oplus x_{0}x_{1}x_{3}x_{6} \oplus x_{0}x_{1}x_{3}x_{6}x_{7} \oplus x_{0}x_{1}x_{3}x_{5}x_{6}x_{7} \oplus x_{0}x_{1}x_{3}x_{4}x_{5}x_{7} \oplus x_{0}x_{1}x_{2}x_{6}x_{7} \oplus x_{0}x_{1}x_{2}x_{5} \oplus x_{0}x_{1}x_{2}x_{5}x_{7} \oplus x_{0}x_{1}x_{2}x_{4} \oplus x_{0}x_{1}x_{2}x_{4}x_{6} \oplus x_{0}x_{1}x_{2}x_{4}x_{5} \oplus x_{0}x_{1}x_{2}x_{4}x_{5}x_{7} \oplus x_{0}x_{1}x_{2}x_{4}x_{5}x_{6} \oplus x_{0}x_{1}x_{2}x_{3} \oplus x_{0}x_{1}x_{2}x_{3}x_{7} \oplus x_{0}x_{1}x_{2}x_{3}x_{5}x_{7} \oplus x_{0}x_{1}x_{2}x_{3}x_{5}x_{6} \oplus x_{0}x_{1}x_{2}x_{3}x_{5}x_{6}x_{7} \oplus x_{0}x_{1}x_{2}x_{3}x_{4}x_{5}$
	}}

	{\footnotesize{$invShiftRows(b_0) = x_{0}$}}

	{\footnotesize{$invMixColumns(b_0) = x_{3} \oplus x_{2} \oplus x_{1} \oplus x_{11} \oplus x_{9} \oplus x_{8} \oplus x_{19} \oplus x_{18} \oplus x_{16} \oplus x_{27} \oplus x_{24}$}}
	\end{flushleft}
	\caption{Boolean equations of deciphering functions for the bit $b_{0}$}
	\label{fig:threefunctions}
\end{figure}

\subsubsection{Solution for the key expansion function}

The key expansion function is the same for both ciphering and deciphering process. Boolean equations we built previously are reusable.

\subsubsection{Global solution}

We have now a Boolean equation for each of \texttt{Inv\-Sub\-Bytes} $ISB()$, \texttt{Inv\-Shift\-Rows} $ISR()$ and \texttt{Inv\-Mix\-Columns} $IMC()$ functions. However, unlike the arrangement of intermediate rounds of the encryption process, these three functions are not combined among them. Indeed, the function \texttt{AddRoundKey} no longer occurs at the end of the round but sits between \texttt{Inv\-Sub\-Bytes} and \texttt{Inv\-Mix\-Columns} functions.

Thus, for a block $B = (b_1, \cdots, b_{128})$ and a key $K = (k_1, \cdots, k_{128})$ as input of the round, the block $B' = (b'_1, \cdots, b'_{128})$ as output is such that: $$B' = IMC(ISB\circ ISR(B) \oplus AD(K))$$

To reduce the Boolean equations, we will not therefore be able to combine the equations of \texttt{Inv\-Sub\-Bytes} and \texttt{Inv\-Shift\-Rows}. As before, to achieve this we just have to replace each input variable of a function with its output value of the previous function using the following equation: $$b'_i = ISB( ISR(b_i) ) \quad\forall i \in (1, \cdots, 128)$$

In python language, the round generation function is written according to the following code \citelst{lst:invround}.

\begin{lstlisting}[style=python, caption={The equation for calculating a decryption round function},label=lst:invround]
def writeRoundDec(numRound, equaSB, equaSR):
	printColor('## Round %s' % numRound, GREEN)
	resultSR = []
	for i in xrange(blockSize):
		equaSR[i] = equaSR[i].split('_')
		resultSR.append(equaSB[int(equaSR[i][1])])
	binMon = generateBinaryMonomes(resultSR)
	return resultSR
\end{lstlisting}

As for the encryption process, we can now describe under the form of Boolean equations the full process of the AES decryption. The function in python language computing this process is given in listing \ref{lst:deciphering} page~\pageref{lst:deciphering}.

\begin{lstlisting}[style=python, caption={Calculation of Boolean functions of the AES decryption process},label=lst:deciphering]
def generateDecFullFiles():
	printColor('## Deciphering process', YELLOW)
	createAESFiles('dec')
	addRoundKey(10, 'dec')
	writeRoundDec(9, invSubBytes(), invShiftRows())
	addRoundKey(9, 'dec')
	writeInvMixColumns(9)
	writeRoundDec(8, invSubBytes(), invShiftRows())
	addRoundKey(8, 'dec')
	writeInvMixColumns(8)
	writeRoundDec(7, invSubBytes(), invShiftRows())
	addRoundKey(7, 'dec')
	writeInvMixColumns(7)
	writeRoundDec(6, invSubBytes(), invShiftRows())
	addRoundKey(6, 'dec')
	writeInvMixColumns(6)
	writeRoundDec(5, invSubBytes(), invShiftRows())
	addRoundKey(5, 'dec')
	writeInvMixColumns(5)
	writeRoundDec(4, invSubBytes(), invShiftRows())
	addRoundKey(4, 'dec')
	writeInvMixColumns(4)
	writeRoundDec(3, invSubBytes(), invShiftRows())
	addRoundKey(3, 'dec')
	writeInvMixColumns(3)
	writeRoundDec(2, invSubBytes(), invShiftRows())
	addRoundKey(2, 'dec')
	writeInvMixColumns(2)
	writeRoundDec(1, invSubBytes(), invShiftRows())
	addRoundKey(1, 'dec')
	writeInvMixColumns(1)
	writeRoundDec(0, invSubBytes(), invShiftRows())
	addRoundKey(0, 'dec')
	writeEndFlag('dec')
	printColor('## Files generated', YELLOW)
\end{lstlisting}

\subsection{Implementation and proof}

We now have two systems of Boolean equations corresponding to the encryption process and decryption of AES. These two systems each have:

\begin{itemize}
	\item 128 equations, one for each bit block;
	\item 1280 variables for the input block;
	\item 1280 variables for the key.
\end{itemize}

Concerning the variables of keys, the fact that we have a Boolean equation by round key involve that we have a set of 128 new variables at each round that is 1280 variables for the AES-128. Each of the variables of the $n$ round key being described in terms of variables of the $n-1$ round key. Consequently and due to the \texttt{XOR} bitwise operation between the round key and the bits resulting from the round function, we are obliged to insert a new set of 128 variables to describe the block transformation at each round.

Finally we described the AES encryption and decryption process in the form of two systems of Boolean equations with 128 equations and 2560 variables.

This mechanism allows us then to describe all of the AES encryption process in the form of files using the same representation as described above. So we have 128 files, one by bit of block. In these files, each line describes a monomial and the transition from one line to the next is done by the \texttt{XOR} operation.

To implement this mechanism of the description of the AES encryption algorithm and generate the 128 files, we have developed and used a python script based on that described earlier in our presentation of AES\footnote{The source file is available at the link \url{https://github.com/archoad/PythonAES}. This program requires a working Python environment it is cross-platform and does not use specific libraries.}.

The main program, \texttt{aes\_equa.py}, offers the possibility of one hand to generate the files for AES ciphering and deciphering functions with the \texttt{generateEncFullFiles()} and \texttt{generateDecFullFiles()} functions and on the other hand, to control that the encryption and the decryption obtained from files is consistent.

Thus, the functions \texttt{controlEncFullFiles()} and \texttt{controlDecFullFiles} performs respectively the encryption and the decryption from the previously generated files. The function \texttt{controlEncFullFiles()} takes as input a block of 128 bits of plain text and a 128-bit block of key while the function \texttt{control\-DecFullFiles()} takes as input a block of 128 bits of cipher text and a a 128-bit block of key. The selected blocks are those provided as test vectors in Appendix B of FIPS 197 \cite{fips197}. The obtained results correspond to those provided in the FIPS: files we generated well represent the AES encryption and decryption algorithm.

\subsubsection{Results obtained from the ciphering process}

The result obtained by the function \texttt{generateEncFullFiles()} is shown in figure~\ref{lst:generate_encfiles} page~\pageref{lst:generate_encfiles} and the result obtained by the \texttt{controlEncFullFiles()} is shown in the listing~\ref{lst:control_encfiles} page~\pageref{lst:control_encfiles}. The control function \texttt{controlEncFullFiles()} injects in the Boolean functions the 128 initial variables corresponding to the clear text block and the 1280 variables corresponding to the key blocks of each round.

\begin{figure}[h]
\centering
\begin{subfigure}{0.38\textwidth}
\begin{lstlisting}[style=bash]
./aes_equa.py
## Ciphering process
## Create directory AES_files
## AddRoundKey0
## Round0
## AddRoundKey1
## Round1
## AddRoundKey2
## Round2
## AddRoundKey3
## Round3
## AddRoundKey4
## Round4
## AddRoundKey5
## Round5
## AddRoundKey6
## Round6
## AddRoundKey7
## Round7
## AddRoundKey8
## Round8
## AddRoundKey9
## Round9
## AddRoundKey10
## Files generated
\end{lstlisting}
\caption{Result of the files creation program for encryption}
\label{lst:generate_encfiles}
\end{subfigure}
\begin{subfigure}{0.58\textwidth}
\begin{lstlisting}[style=bash]
 ./aes_equa.py
## Clear block 00112233445566778899aabbccddeeff
## Key block 000102030405060708090a0b0c0d0e0f
## addRoundKey0
00102030405060708090a0b0c0d0e0f0 32
## Round0
5f72641557f5bc92f7be3b291db9f91a 32
## addRoundKey1
89d810e8855ace682d1843d8cb128fe4 32
## Round1
ff87968431d86a51645151fa773ad009 32
## addRoundKey2
4915598f55e5d7a0daca94fa1f0a63f7 32
## Round2
4c9c1e66f771f0762c3f868e534df256 32
## addRoundKey3
fa636a2825b339c940668a3157244d17 32
## Round3
6385b79ffc538df997be478e7547d691 32
## addRoundKey4
247240236966b3fa6ed2753288425b6c 32
## Round4
f4bcd45432e554d075f1d6c51dd03b3c 32
## addRoundKey5
c81677bc9b7ac93b25027992b0261996 32
## Round5
9816ee7400f87f556b2c049c8e5ad036 32
## addRoundKey6
c62fe109f75eedc3cc79395d84f9cf5d 32
## Round6
c57e1c159a9bd286f05f4be098c63439 32
## addRoundKey7
d1876c0f79c4300ab45594add66ff41f 32
## Round7
baa03de7a1f9b56ed5512cba5f414d23 32
## addRoundKey8
fde3bad205e5d0d73547964ef1fe37f1 32
## Round8
e9f74eec023020f61bf2ccf2353c21c7 32
## addRoundKey9
bd6e7c3df2b5779e0b61216e8b10b689 32
## Round9
7ad5fda789ef4e272bca100b3d9ff59f 32
## addRoundKey10
69c4e0d86a7b0430d8cdb78070b4c55a 32
69c4e0d86a7b0430d8cdb78070b4c55a (FIPS result)
\end{lstlisting}
\caption{Result of the files control program for encryption}
\label{lst:control_encfiles}
\end{subfigure}
\end{figure}

\subsubsection{Results obtained from the deciphering process}

According to the same principle as for Boolean functions of encryption, the result obtained by the function \texttt{generateDecFullFiles()} is shown in the listing~\ref{lst:generate_decfiles} page~\pageref{lst:generate_decfiles} and the obtained result from the \texttt{controlDecFullFiles()} function is shown in the listing~\ref{lst:control_decfiles} page~\pageref{lst:control_decfiles}.

\begin{figure}[h]
\centering
\begin{subfigure}{0.38\textwidth}
\begin{lstlisting}[style=bash]
./aes_equa.py
## Deciphering process
## Create directory AES_files
## AddRoundKey10
## Round 9
## AddRoundKey9
## InvMixColumns 9
## Round 8
## AddRoundKey8
## InvMixColumns 8
## Round 7
## AddRoundKey7
## InvMixColumns 7
## Round 6
## AddRoundKey6
## InvMixColumns 6
## Round 5
## AddRoundKey5
## InvMixColumns 5
## Round 4
## AddRoundKey4
## InvMixColumns 4
## Round 3
## AddRoundKey3
## InvMixColumns 3
## Round 2
## AddRoundKey2
## InvMixColumns 2
## Round 1
## AddRoundKey1
## InvMixColumns 1
## Round 0
## AddRoundKey0
## Files generated
\end{lstlisting}
\caption{Result of the file creation program for decryption}
\label{lst:generate_decfiles}
\end{subfigure}
\begin{subfigure}{0.58\textwidth}
\begin{lstlisting}[style=bash]
./aes_equa.py
## Cipher block 69c4e0d86a7b0430d8cdb78070b4c55a
## Key block 000102030405060708090a0b0c0d0e0f
## addRoundKey10
7ad5fda789ef4e272bca100b3d9ff59f 32
## Round9
bd6e7c3df2b5779e0b61216e8b10b689 32
## addRoundKey9
e9f74eec023020f61bf2ccf2353c21c7 32
## invMixColumns9
54d990a16ba09ab596bbf40ea111702f 32
## Round8
fde3bad205e5d0d73547964ef1fe37f1 32
## addRoundKey8
baa03de7a1f9b56ed5512cba5f414d23 32
## invMixColumns8
3e1c22c0b6fcbf768da85067f6170495 32
## Round7
...
## Round3
fa636a2825b339c940668a3157244d17 32
## addRoundKey3
4c9c1e66f771f0762c3f868e534df256 32
## invMixColumns3
3bd92268fc74fb735767cbe0c0590e2d 32
## Round2
4915598f55e5d7a0daca94fa1f0a63f7 32
## addRoundKey2
ff87968431d86a51645151fa773ad009 32
## invMixColumns2
a7be1a6997ad739bd8c9ca451f618b61 32
## Round1
89d810e8855ace682d1843d8cb128fe4 32
## addRoundKey1
5f72641557f5bc92f7be3b291db9f91a 32
## invMixColumns1
6353e08c0960e104cd70b751bacad0e7 32
## Round0
00102030405060708090a0b0c0d0e0f0 32
## addRoundKey0
00112233445566778899aabbccddeeff 32
00112233445566778899aabbccddeeff (FIPS result)
\end{lstlisting}
\caption{Result of the files control program for decryption}
\label{lst:control_decfiles}
\end{subfigure}
\end{figure}

In both cases, encryption and decryption, the results we obtain by using our files to cipher and to decipher blocks are conform to those described in the FIPS 197. So our Boolean equation system describing the AES algorithm is right.

\section{Conclusion}

After presenting briefly the Boolean algebra, Boolean functions and two of their presentations, we have developed a process that allows us to translate the AES encryption and decryption algorithms in Boolean functions. Then we defined a mode of representation of these Boolean functions in the form of computer files. Finally, we have developed a program to implement this process and to check that the expected results are consistent with those provided in the FIPS.

In the end, we got a two new systems of Boolean equations, the first one describing the entire ciphering process while the second describes the entire deciphering process of the \textsl{Advanced Encryption Standard} and each one including 128 equations and $(128 \times 10) + (128 \times 10) = 2560$ variables.

The next step could be to search, through statistical and combinatorial analysis, new ways to cryptanalyse the AES. Either by finding a solution to resolve our equations system either by using statistical bias exploitable with this system.

\end{document}